\documentclass[%
 reprint,
nofootinbib,
 amsmath,amssymb,
 aps,
floatfix
]{revtex4-2}
\usepackage{hyperref}
\usepackage{graphicx}% Include figure files
\usepackage{dcolumn}% Align table columns on decimal point
\usepackage{bm}% bold math
\usepackage[super]{nth}
\usepackage{aasmacros}
\usepackage{xcolor}

\begin{document}

\preprint{APS/123-QED}

\title{Nonlinear 3D Cosmic Web Simulation with Heavy-Tailed Generative Adversarial Networks}

\author{Richard M. Feder}%
\email{rfederst@caltech.edu}
\affiliation{%
 California Institute of Technology Division of Physics, Math, and Astronomy, 1200 East California Boulevard, Pasadena, CA 91125, USA
}%
\author{Philippe Berger}
\affiliation{Jet Propulsion Laboratory, California Institute of Technology, Pasadena, California 91109, USA}%Lines break automatically or can be forced with 
\author{George Stein}
\affiliation{
 Berkeley Center for Cosmological Physics, University of California, Berkeley, CA 94720, USA
}%

\date{\today}% It is always \today, today,
             %  but any date may be explicitly specified

\begin{abstract}

 Fast and accurate simulations of the non-linear evolution of the cosmic density field are a major component of many cosmological analyses, but the computational time and storage required to run them can be exceedingly large. For this reason, we use generative adversarial networks (GANs) to learn a compressed representation of the 3D matter density field that is fast and easy to sample, and for the first time show that GANs are capable of generating samples at the level of accuracy of other conventional methods. Using sub-volumes from a suite of GADGET-2 N-body simulations, we demonstrate that a deep-convolutional GAN can generate samples that capture both large- and small-scale features of the matter density field, as validated through a variety of $n$-point statistics. The use of a data scaling that preserves high-density features and a heavy-tailed latent space prior allow us to obtain state of the art results for fast 3D cosmic web generation. In particular, the mean power spectra from generated samples agree to within $5\%$ up to $k=3$ and within $10\%$ for $k\leq 5$ when compared with N-body simulations, and similar accuracy is obtained for a variety of bispectra. By modeling the latent space with a heavy-tailed prior rather than a standard Gaussian, we better capture sample variance in the high-density voxel PDF and reduce errors in power spectrum and bispectrum covariance on all scales. Furthermore, we show that a conditional GAN can smoothly interpolate between samples conditioned on redshift. Deep generative models, such as the ones described in this work, provide great promise as fast, low-memory, high-fidelity forward models of large-scale structure. 

\end{abstract}

%\keywords{Suggested keywords}%Use showkeys class option if keyword
                              %display desired
\maketitle

%\tableofcontents

\section{\label{sec:level1}Introduction}

Modeling the evolution of the large-scale structure of the Universe is a crucial part of many cosmological analyses. The initial linear growth of structures becomes highly non-linear at later times and smaller scales, such that analytical approaches are unable to accurately predict observations in those regimes. For this reason, numerical simulations \cite{cosmosims} of a representative volume of the universe are required to bridge the gap between the early universe and present day observations. 

N-body simulations, which numerically solve the Poisson equation, are a standard operating procedure for modeling large-scale structure. These methods are exact in the sense that they directly simulate the gravitational evolution of some set of collisionless dark matter ``particles'', but the required mass and time resolution to make model predictions at scale is computationally expensive and requires significant data storage. Next-generation cosmological inferences will require thousands of simulations on the largest scales in order to constrain models like $\Lambda$CDM at the sub-percent level \citep{blot_realizations}, a task that is currently intractable to all but the largest future computational grants when using standard N-body methods. Another consideration is memory storage. At the moment, storing thousands of full scale N-body simulation snapshots is only feasible for petabyte-scale data storage facilities. Even with access to such resources, the analysis workflow may be cumbersome if one has to repeatedly transfer simulations between a cluster and local machine.

The need for fast and accurate simulations at a cheaper computational cost has led to the development of many ``approximate'' methods of cosmic structure simulation \cite{cola, icecola, scola, fastpm, pinocchio, manera, apt, halogen, mpp}, which implement various physical approximations of gravity to decrease the simulation run-time and/or memory requirement while attempting to maintain a high degree of accuracy. These methods have shown to be very useful, but still require a non-negligible amount of computation time, and the accuracy of their results has room to be improved upon. 

One may then wonder if moving past physical approximations of gravity, and instead allowing for arbitrary non-linear mappings from initial conditions (or some other prior distribution) to present day-observables, could provide more accurate simulations at a smaller computational cost. This application of machine learning (ML) techniques to aid in the efficient modeling of large scale structure has not gone un-investigated, and there has recently been great interest in the use of ML across the field\footnote{Comprehensive list at \href{https://github.com/georgestein/ml-in-cosmology}{github.com/georgestein/ml-in-cosmology}} \cite{ml-in-cosmo}. Image-to-image mapping techniques that transform initial conditions to the final structures \cite{halonet, d3m, halonetv2} or augment dark matter-only simulation results with various physical effects not present in the simulation \cite{nnreconstruction, neutrino-unet, dmfeedback, painting1, painting2, dm2gal} have shown promise. However, these still require the generation of initial conditions, or the full simulation output, respectively. A fully generative model that does not require these intermediate steps is desirable.

In particular, generative adversarial networks (GANs) have garnered attention in recent years due to their ability to generate fast realizations of large scale structure. GANs are one class of deep generative modeling techniques that seeks to learn a probability distribution over an arbitrary data distribution, doing so through a two-player adversarial optimization scheme \cite{goodfellow} (see Section \ref{sec:gan} for more details).   

To date, there have been a number of successful implementations of cosmological modeling tasks using GANs. They have been used to generate accurate weak lensing convergence maps as evaluated through the power spectrum and a variety of Minkowski functionals \cite{weak_lens_gan}, to model 2D slices of the cosmic density field from 3D N-body simulations \cite{cosmicweb_gan}, and to generate 3D realizations of neutral hydrogen \cite{higan}. These methods learn a compressed representation of the data distribution and sample from the corresponding lower-dimensional latent space to produce new samples.

There are also a number of in-painting and super-resolution methods that do not involve the determination of a compressed data representation. One promising method is \cite{super_res_nbody}, in which a super-resolution GAN and deep physical model are used to map low resolution, 3D N-body simulations to their high resolution counterparts. This is similar in concept to \cite{painting1} and \cite{painting2}, which used GANs to map dark matter density fields to corresponding halo number count maps and hydrodynamical quantities, respectively. \cite{scale_up_gan1} propose a super-resolution scheme for generating large-scale realizations of the matter density field hierarchically, treating the scalability problem separately from sample accuracy.

Three-dimensional N-body data sets are challenging to model because each sample realization contains many degrees of freedom. In addition, the voxel PDF for each realization has large tails, with matter densities spanning several orders of magnitude. Natural image data sets, a common focus of GAN applications, also contain tails in pixel space \citep{natural_image_stats}, however the application in cosmology is unique because the tails of the voxel PDF often have a disproportionate impact on the relevant summary statistics.

This exploratory work identifies a number of factors important for training GANs that, from a compressed representation, produce samples that are both accurate and preserve the sample variance of the data distribution. In a controlled setting, we demonstrate the need for data scalings that preserve information in the high-density regime. Furthermore, we find that heavy-tailed distributions are more suitable priors for modeling the latent space, significantly improving the accuracy of our generated covariance and correlation matrices. Through these and other considerations, we obtain state of the art results for 3D cosmic web generation as reflected through the matter density power spectrum, a range of bispectra, and other cosmological summary statistics.

The report is structured as follows: in \S 2 we introduce the statistical and optimization framework of GANs. We describe the GADGET-2 data set and training procedure in \S 3 and outline the modeling choices specific to this work. The corresponding results for single redshift and redshift-interpolated models are presented in \S 4. Finally, we discuss our results in \S 5, concluding with a discussion of future directions and applications.

\section{Generative Adversarial Networks} \label{sec:gan}

Generative adversarial networks (GANs), originally introduced in \cite{goodfellow}, are a class of unsupervised models that learn an underlying probability distribution, which we denote $p_{data}(x)$, through adversarial training. In this paradigm, a generator $G$ attempts to produce samples that appear real to a discriminator $D$. The discriminator is tasked with distinguishing between real training samples and synthetic ones.  

\begin{figure*}[t]
    \centering
    \includegraphics[width=0.65\linewidth]{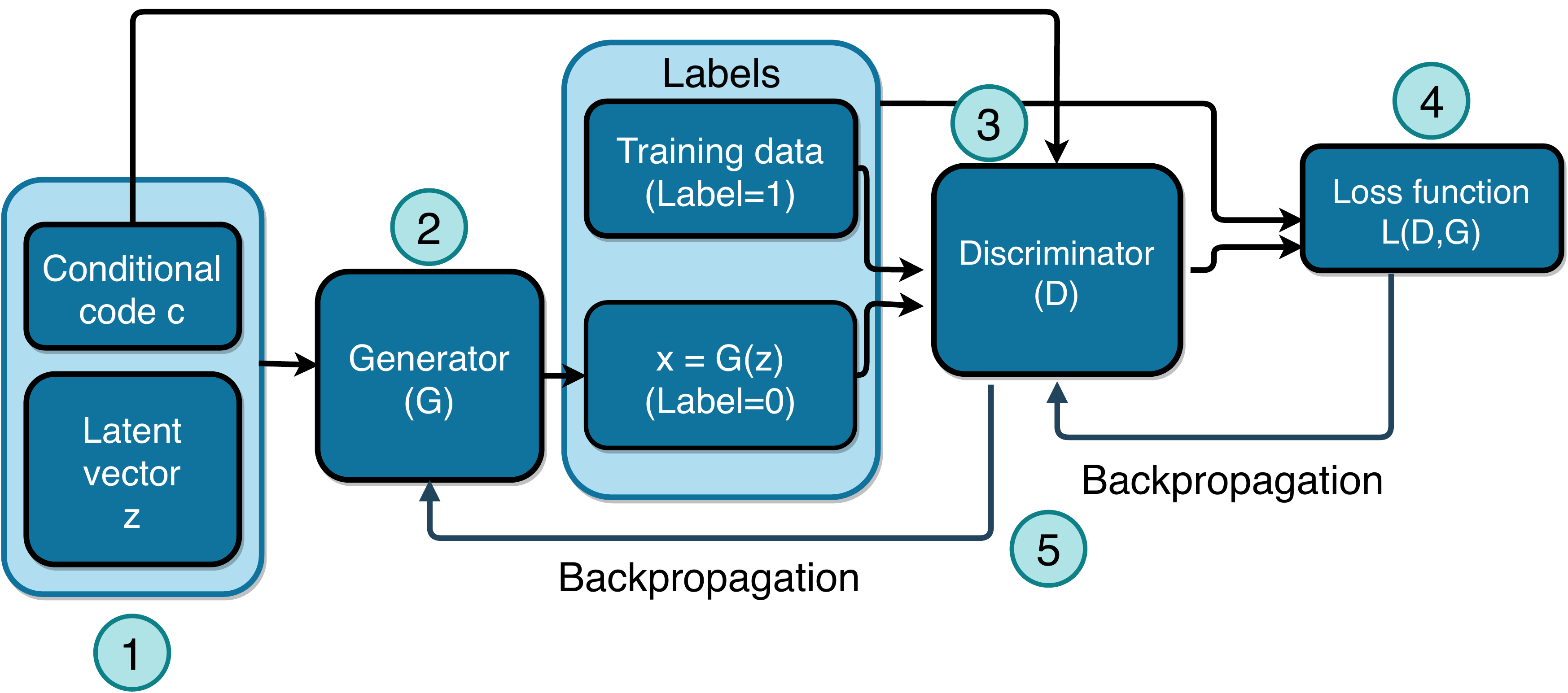}
    \caption{Schematic for generative adversarial networks. As described in Section \ref{sec:gan}, a latent vector is drawn from some prior (1) and used as input for the generator network (2) which outputs a sample in the data space. A batch of these samples, along with a batch of training set samples, are passed to the discriminator (3). The discriminator network outputs a set of scores representing how much it believes the samples come from the data distribution, and these scores are compared with ground truth labels to calculate a loss (4). This loss is then used to update the parameters of both the generator and discriminator network through backpropagation (5).}
    \label{fig:GAN_setup}
\end{figure*}

The standard training process, shown schematically in Figure \ref{fig:GAN_setup}, goes as follows:
\begin{enumerate}
    \item The input is either a random latent vector $\mathbf{z}$ (typically drawn from a unit-normal distribution $\mathcal{N}\left(0, \mathbf{I}\right)$), or some structured input. 
    \item The generator $G$, parameterized by a neural network, takes the latent vector $\mathbf{z}$ and outputs a sample $G(\mathbf{z})$. 
    \item The discriminator $D$, also parameterized by a neural network, takes real samples $\mathbf{x}$ and generated samples $G(\mathbf{z})$ as input, and outputs scores $D(\mathbf{x})$ and $D(G(\mathbf{z}))$, respectively. Each score represents the degree to which the discriminator believes a given sample comes from the data distribution $p_{data}(\mathbf{x})$. When the score is scaled to the range $[0, 1]$, it is sometimes interpreted as an implicit likelihood of the data given the discriminative model, $p(\mathbf{x}|D)$. 
    
    \item The discriminator predictions are compared to the ground truth labels, from which a loss $\mathcal{L}(D,G)$ is computed.
    \item The loss is used to update the weights of $G$ and $D$ through backpropagation, flowing through the discriminator and then through the generator.
    \item Steps 1-5 are repeated, looping through the training set over several epochs.
\end{enumerate}

Formally, GAN training is a min-max problem, i.e. the generator and discriminator compete to minimize and maximize a loss function, respectively:
\begin{multline}
    \min_G \max_D \mathcal{L}(G,D) = \mathbb{E}_{\mathbf{x}\sim p_{data}(\mathbf{x})}\left[\log(D(\mathbf{x}))\right] \\ + \mathbb{E}_{\mathbf{z}\sim p_g(\mathbf{z})}\left[\log(1 - D(G(\mathbf{z})))\right] 
\label{gan_objective}
\end{multline}
The first expectation on the right hand side updates the discriminator, while the second term  updates both the discriminator and the generator. By training both networks simultaneously, the generator gradually learns to produce samples from the data distribution. As described in \cite{goodfellow}, the optimal generator and discriminator reach a Nash equilibrium, where neither network can make unilateral improvement over the other. That work also demonstrated that, for an optimal discriminator and certain choices of loss function, training the generator is equivalent to reducing the Jensen-Shannon divergence between the generated distribution $p_g(z)$ and the empirical distribution $p_{data}(x)$. The Jensen-Shannon divergence is the symmetrized version of the Kullback-Leibler divergence:
\begin{equation}
    \mathbb{D}_{JS}(p_g(\mathbf{x})|p_{data}(\mathbf{x})) = \frac{1}{2}\mathbb{D}_{KL}(p_g||M)+\frac{1}{2}\mathbb{D}_{KL}(p_{data}||M)
\end{equation}
where
\begin{equation}
M = \frac{p_{data}(\mathbf{x})+p_g(\mathbf{z})}{2}.
\end{equation}

A common assumption when using GANs for modeling is that while the observed data may be high-dimensional, the underlying data distribution exists on a lower-dimensional manifold. In this spirit, generator outputs $G(\mathbf{z})$ may be interpreted as samples drawn from some underlying manifold $\mathcal{G}$, for which $\dim\mathcal{G} = |\mathbf{z}|$ and $\text{Cov}(\mathcal{G}) = G$, where $G$ is the neural network. In many applications, $\mathbf{z}$ is chosen to have dimension $\mathcal{O}(100)$, though other techniques like in-painting and super-resolution may use higher-dimensional inputs. One advantage of learning a compressed representation of the data is that generating high-dimensional samples in the data space reduces to sampling from a simple, lower-dimensional latent space and performing a forward pass through the generator network $G$. This also means that sample generation is fast, since the only computation needed after a draw from the latent distribution is a forward pass through the network.

\subsection{Conditional GANs}
There are cases where one would like to incorporate additional information relevant to the data generation process. The conditional GAN (cGAN) is a natural extension of the GAN that addresses these cases \citep{cgan}. The training problem is a modified version of the original objective:
\begin{multline}
    \min_G \max_D \mathcal{L}(G,D) = \mathbb{E}_{\mathbf{x}\sim p_{data}(\mathbf{x})}\left[\log(D(\mathbf{x}|\mathbf{c}))\right]  \\ + \mathbb{E}_{\mathbf{z}\sim p_g(\mathbf{z})}\left[\log(1 - D(G(\mathbf{z}|\mathbf{c})))\right]. 
\end{multline}
Here, $\mathbf{c}$ denotes conditional information, which can be appended both to the latent vector of the generator and to the discriminator as an extra feature. A properly trained conditional generator is then able to interpolate within the conditional distribution $p(\mathbf{x}|\mathbf{c})$. If the marginal distribution $p(\mathbf{c})$ is known, then training the cGAN results in a model that can generate samples from the joint distribution $p(\mathbf{x},\mathbf{c}) = p(\mathbf{c})p(\mathbf{x}|\mathbf{c})$. For example, in the case of the standard MNIST data set of handwritten digits\footnote{\url{http://yann.lecun.com/exdb/mnist/}}, an obvious conditional distribution is $c \sim \text{Cat}(k)$, where $k\in \lbrace 0,1,\hdots,9\rbrace$. There is nothing that specifically constrains the form or dimensionality of the conditional information, so long as it can be passed through layers of the generator and discriminator. 

The theoretical guarantees of GANs are compromised by a number of factors in practice. Because optimization is performed over network parameters and not over probability densities, convergence to the global optimum where $p_g = p_{data}$ may be complicated by critical points and/or inefficient training. Furthermore, unstable training dynamics can lead one of the networks to dominate the other such that gradients are suppressed and learning stops prematurely. When the generator dominates the discriminator, the pathology is referred to as ``mode collapse", in which the generator learns to model some small subset of the data distribution. Despite these practicalities, GANs with reasonable model capacity and sufficient training data can be trained successfully and produce high-fidelity samples from the underlying data distribution \citep{stylegan, biggan}.

\section{Data set and training procedure}

\subsection{N-body simulations with GADGET-2}

While many types of cosmological data have non-linear properties, we choose to train a GAN to model the dark matter component of the ``cosmic web" \citep{bond}. The cosmic web is a combination of baryonic and dark matter that evolves with the expansion of the universe and eventually clusters, forming the overdense seeds of galaxy and star formation. Due to properties of gravitational collapse in an expanding universe, the cosmic web is comprised of many coherent structures, such as one-dimensional filaments and two-dimensional sheets.

Our training set comes from a suite of thirty-two dark matter only simulations made with the software GADGET-2 \footnote{\url{https://wwwmpa.mpa-garching.mpg.de/gadget/}}. Each simulation is initialized with a unique seed, using the cosmological parameters that match the best-fitting results of the
WMAP+BOSS DR9 analysis \cite{cosmoparams}. The initial conditions are derived using second-order Lagrangian perturbation theory (2LPT) at a starting redshift of $z=49$.
Each simulation volume has 512$^3$ particles in a (512 Mpc)$^3$ volume that are evolved to the present day, with snapshots written to disk at $z=\lbrace 3.0, 2.0, 1.5, 1.0, 0.5, 0.25, 0.0 \rbrace$, after which they are placed on a 3D grid with 512 resolution elements per side. After gridding, the GADGET-2 density field is given in units of $10^{10} M_{\odot}/\text{cell}$, and we will refer to this normalized matter density field as $\widetilde{\rho}(x)$. 

A naive application of GANs to directly generate $512^3$-voxel samples would be challenging because of the high dimension of the data and the availability of training samples. Instead, we train on $(64 \text{ Mpc})^3$ sub-volumes, which gives us a training set with 16384 samples. We further exploit the isotropy of our samples to augment our data set through random rotations and flips, which increases the effective training set size by a factor of $6\text{ (sides)}\times 4 \text{ (90 deg. rotations)} \times 2 \text{ (random flips)} = 48$.

\subsection{Network architecture}
We base our network architecture on \cite{3dgan}, in which the generator upsamples a latent vector $\mathbf{z}$ with 3D transpose convolutional layers and the discriminator downsamples with 3D convolutional layers. This is one example of a deep convolutional GAN (DCGAN, \cite{dcgan}). The parameters and activations used to configure the network are shown in Table \ref{table:net}. 

While a variety of large-scale, sophisticated neural network architectures have been presented in the machine learning literature, we found that (relatively) smaller networks had sufficient capacity to generate high-fidelity samples for our application and were also easier to train. In particular, our results did not change significantly when the dimension of the latent space was increased, when the number of filters was increased, or when extra convolutional layers were added beyond the base architecture.

\begin{table}[h]
\centering
\begin{tabular}{|c|c|}
\hline
Filter sizes & $\lbrace{4,4,4,4,4\rbrace}$ \\
\hline
$n_{filter}$ & $\lbrace{256, 128, 64, 32, 1 \rbrace}^*$ \\
\hline
Strides & $\lbrace{1,2,2,2,2 \rbrace}^*$ \\
\hline
Padding: & $\lbrace{ 0,0,0,0,1 \rbrace}^*$ \\
\hline
Layer Act. &   Leaky ReLU (G), ReLU (D) \\
\hline
Final Act. & Tanh (G), Sigmoid (D) \\
\hline
Latent dimension & 100 \\
\hline
$N_{params}$ & 4,393,920 \\
\hline
\end{tabular}
\caption{Architecture specifications for each layer of the generator network (G) and discriminator network (D). Asterisk signifies that order of layers is reversed in the discriminator.}
\label{table:net}
\end{table}

\subsection{Data scaling}\label{sec:data_scaling}
The evolved matter density field, $\widetilde{\rho}(x)$, exhibits a dynamic range spanning several orders of magnitude. While not strictly prohibited, directly generating samples with values in this range would be difficult for a generator network. In fact, \cite{dcgan} argues that a bounded final activation allows the model to more quickly saturate the range of the training distribution. We use the transformation from \cite{cosmicweb_gan},
\begin{equation}
    c(x) = \frac{2\widetilde{\rho}(x)}{\widetilde{\rho}(x)+\kappa}-1, 
    \label{scale_equation}
\end{equation}
which scales our density fields to the range $[-1, 1)$, suitable for a generator network with a hyperbolic tangent (tanh) final activation.

\begin{figure}[h]
    \centering
    \includegraphics[width=\linewidth]{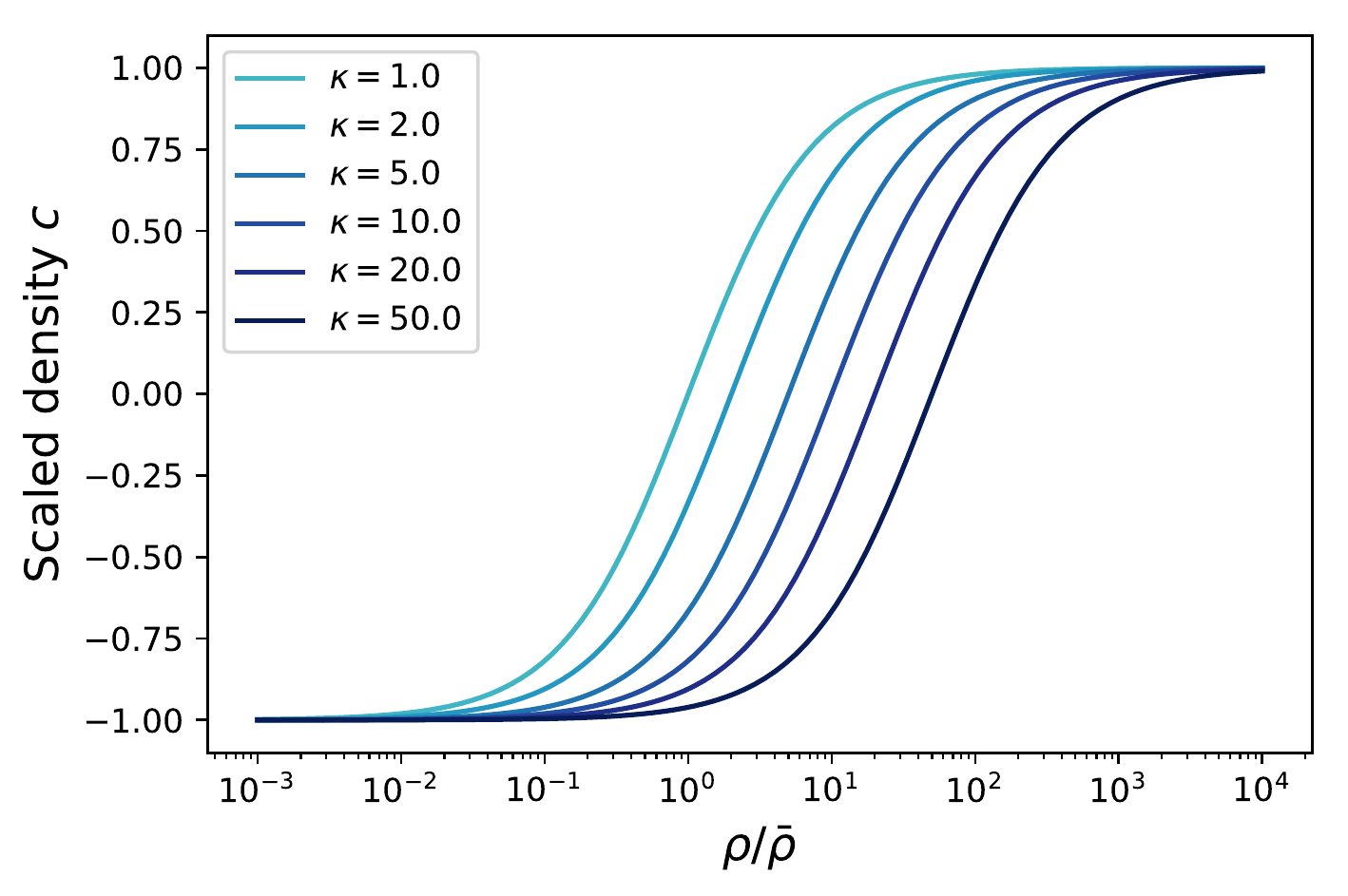}
    \caption{Scaling from Equation \eqref{scale_equation} plotted for different values of $\kappa$. Scalings with small values of $\kappa$ have smaller gradients at high density.}
    \label{fig:data_scaling}
\end{figure}

In \cite{cosmicweb_gan}, the scaling parameter $\kappa$ is set to 4. However, after performing a series of experiments, we choose to adopt a larger value, $\kappa=45$, that better preserves information in the high-density regime and results in more accurate samples. Figure \ref{fig:data_scaling} illustrates how a small variation in scaled density can correspond to a large variation in the final matter density (see Appendix \ref{app:scaling_param} for more detailed analysis). We experimented with the logarithmic scaling used in \cite{higan} as well as other piecewise scalings, but found none were able to capture the statistics of the matter density field as well as Equation \eqref{scale_equation}.

\subsection{Modeling the latent space with a heavy-tailed prior}
The dynamic range of voxel matter densities and of the matter density PDF make N-body simulations a unique data set. While many data sets have non-Gaussian tails in their pixel/voxel distributions, cosmological N-body data sets are particularly challenging to model because the least frequent, high-density voxels have the largest impact on the relevant summary statistics. For the data set considered in this work, the challenge is in capturing the mean voxel distribution function at high density while also capturing its sample variance. 

The standard approach for sample generation with a GAN is to draw a noise vector $\mathbf{z}$ from a Gaussian or uniform distribution. However, there is no strong motivation to use these as priors (beyond ease of sampling) if they do not reflect important features of the data. As shown in Section \ref{sec:results_z0}, it is only when a heavy-tailed prior is used that \emph{both} accuracy and sample variance are captured by our model. To date, and to our knowledge, there are not many instances in the machine learning literature where heavy-tailed distributions are used to model the latent space. \cite{nonparametric_gan} develops non-parametric priors through on-the-shelf optimization tools in order to improve latent space interpolation, though our work is less focused on semantic interpolation than common applications where natural image data sets are modeled. We choose a Student's t-distribution prior for the latent space and tune $\nu$, the number of degrees of freedom in the distribution. The univariate Student's t-distribution has probability density
\begin{equation}
    P(z) = \frac{\Gamma\left(\frac{\nu+1}{2}\right)}{\sqrt{\nu\pi}\Gamma\left(\frac{\nu}{2}\right)}\left(1 + \frac{z^2}{\nu}\right)^{-\frac{\nu+1}{2}}
\end{equation}
where $\Gamma(\cdot)$ is the Gamma function. When $\nu$ is smaller, the Student's t-distribution has larger tails, and in the limit where $\nu \to \infty$, $P(z)$ converges to a Gaussian distribution. Through empirical tests, we find $\nu = 10$ performs best for our data set. We reiterate that our argument for modeling the latent space is not specific to the Student's t-distribution. One might capture tail behavior with a Lorentzian distribution, a Cauchy distribution, or even a mixture of Gaussians. 

\subsection{Training Procedure}

The generator and discriminator were trained for 150 epochs with learning rates of 1 $\times$ 10$^{-4}$ and 5 $\times$ 10$^{-5}$, respectively, using the ADAM optimizer \citep{adam_optimizer}. We chose to train with a batch size of 16, meaning gradient updates to the networks at each iteration are computed based on sixteen real and sixteen generated samples. Within each epoch, each training set mini-batch is chosen randomly, after which a set of random rotations and flips are applied to the samples. Following the training procedure from \cite{3dgan}, the discriminator's weights are only updated in a given batch iteration if its accuracy for that batch falls below 80\%. This is one heuristic to prevent the discriminator from outperforming the generator, and in our case led to more stable training.

We also conducted several experiments with the Wasserstein GAN \cite{wgan} with gradient penalty \cite{wgan_gp} scheme (WGAN-GP). The WGAN-GP trained stably for a range of architectures, optimizers and hyperparameters. While the resulting models were able to capture the mean summary statistics of the data as described in \ref{sec:results_z0}, they struggled to capture sample variance in the generator distribution. Because of this, we chose to use the standard GAN for the remainder of our experiments.

After some training, a GAN may be near its optimum, but if the learning rate is too large then the networks will be limited in their ability to further converge. With this in mind, we use a learning rate scheduler which periodically reduces the learning rate by a factor $\gamma$. We choose to decay the learning rates of both the generator and discriminator by 50\% ($\gamma = 0.5$) after every 1000 gradient updates to the generator. 

\begin{figure}[h]
    \centering
    \includegraphics[width=\linewidth]{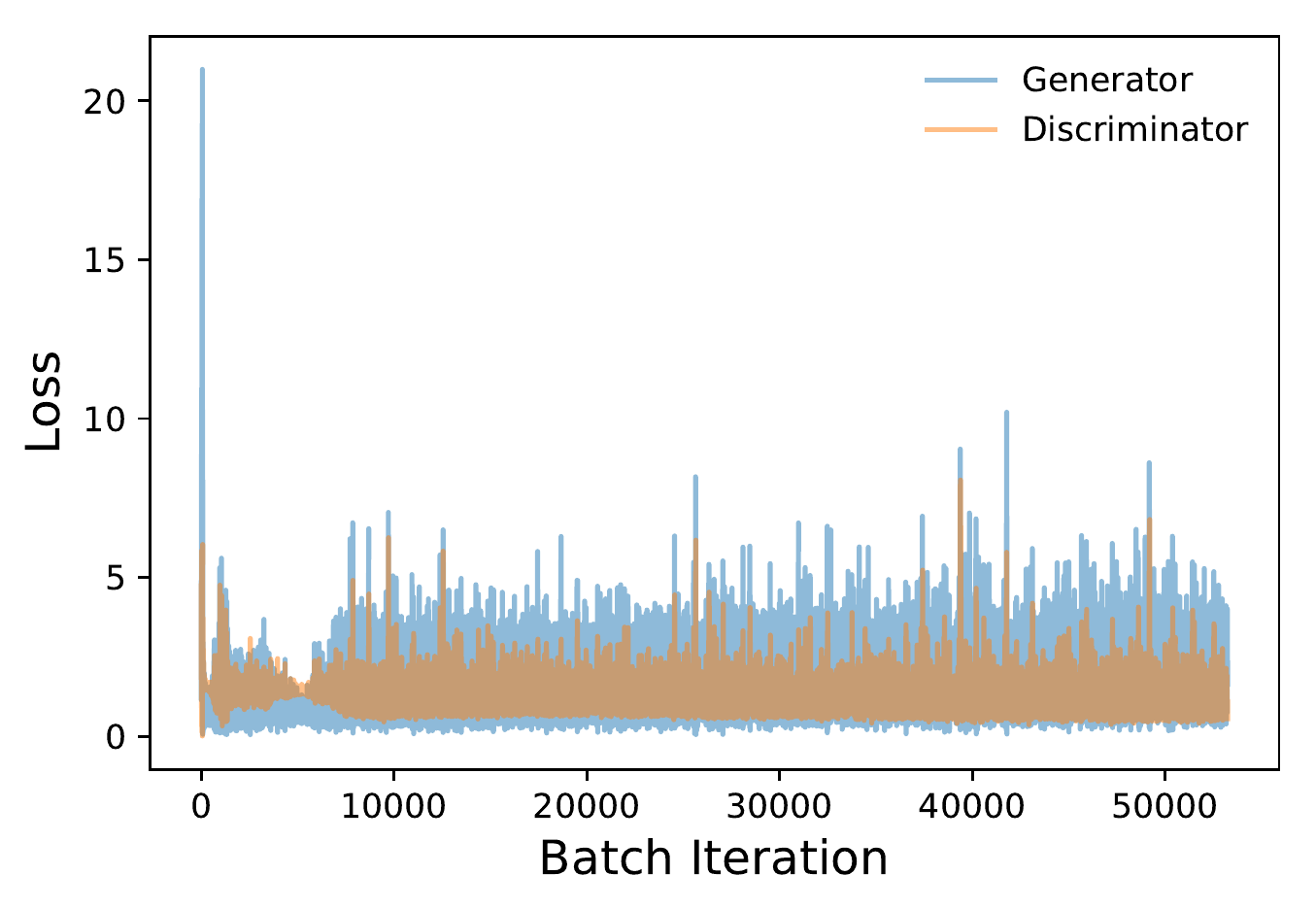}
    \caption{Losses of the generator (blue) and discriminator (orange) as a function of batch iteration for a training run using a 3D DCGAN. The generator loss is computed from the second term in the objective from Equation \eqref{gan_objective}, while the discriminator loss is the sum of the first and second terms. The losses do not indicate when the models have converged, but help verify that the training dynamics of the GAN are controlled.}
    \label{fig:train}
\end{figure}

The losses of the generator and discriminator are plotted as a function of batch iteration in Figure \ref{fig:train}. Despite small fluctuations which come from noisy estimates of the loss computed from mini-batches, training stabilizes relatively quickly. The losses do not indicate when the models have converged, but help verify that the training dynamics of the GAN are controlled. As there are no asymptotic convergence guarantees when training GANs, one is at liberty to choose the model at any point in training when it best captures the desired summary statistics. To optimize our model selection, we save a version of the generator at the end of each epoch and determine for which model the mean voxel PDF and power spectra match the data best. As discussed in Appendix \ref{app:drs}, we find a correspondence between the variance of the latent distribution and the mean power spectral amplitude of generated samples. After training, we use this to further calibrate the model to the training set by perturbing the scale width of the latent distribution, typically at the percent level. 

Computations were done on the Texas Advanced Computing Cluster on a node with 4 $\times$ Nvidia 1080-TI GPUs with 128 GB of RAM, and took roughly nine hours in wall clock time to train. The model and training were implemented using the PyTorch library\footnote{\href{https://pytorch.org/}{https://pytorch.org/}}. We have made the code for this work publicly available on Github\footnote{\href{https://github.com/RichardFeder/gan-work}{https://github.com/RichardFeder/gan-work}}.

\begin{figure*}[t]
\centering

\includegraphics[width=0.3\textwidth]{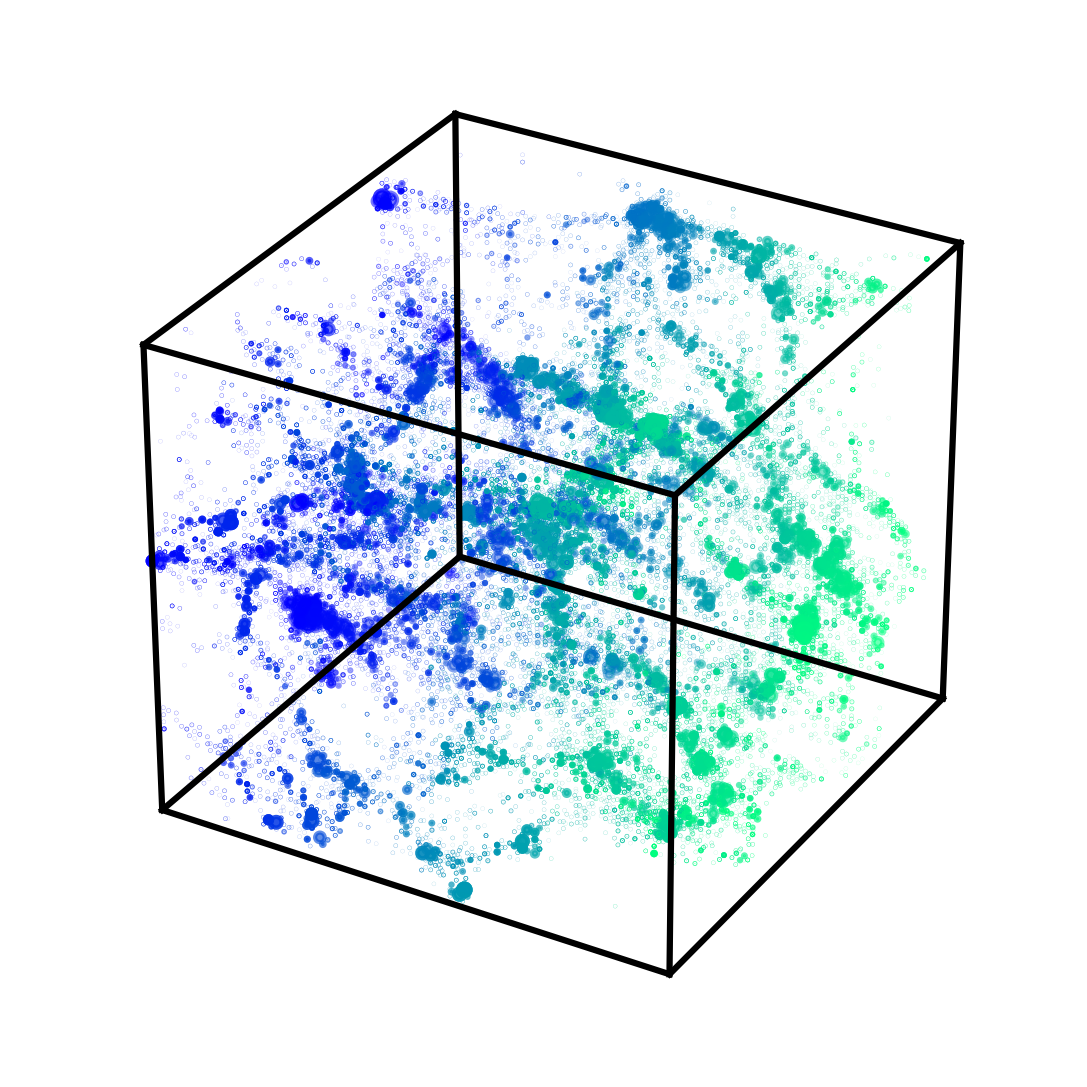}
\includegraphics[width=0.3\textwidth]{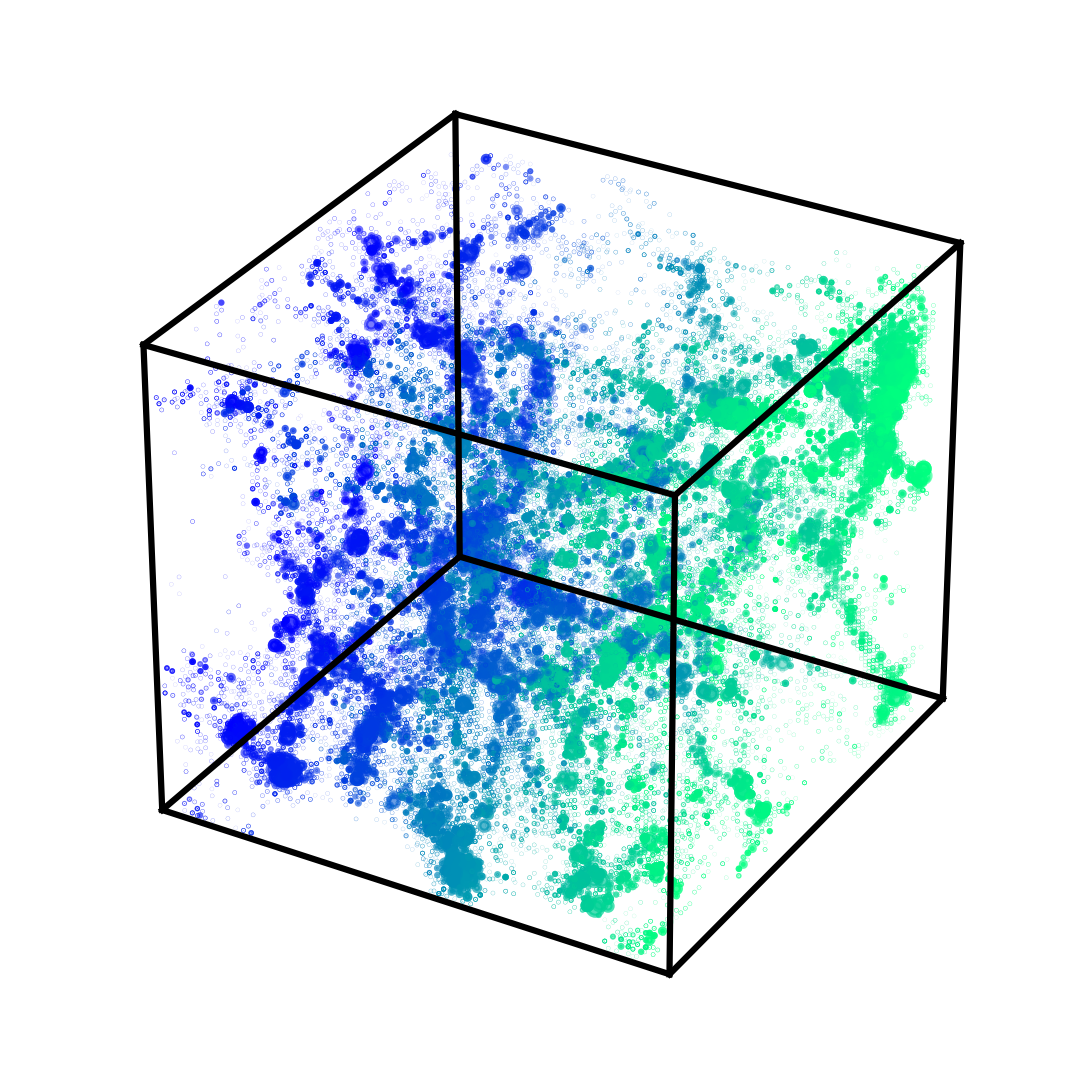}
\includegraphics[width=0.3\textwidth]{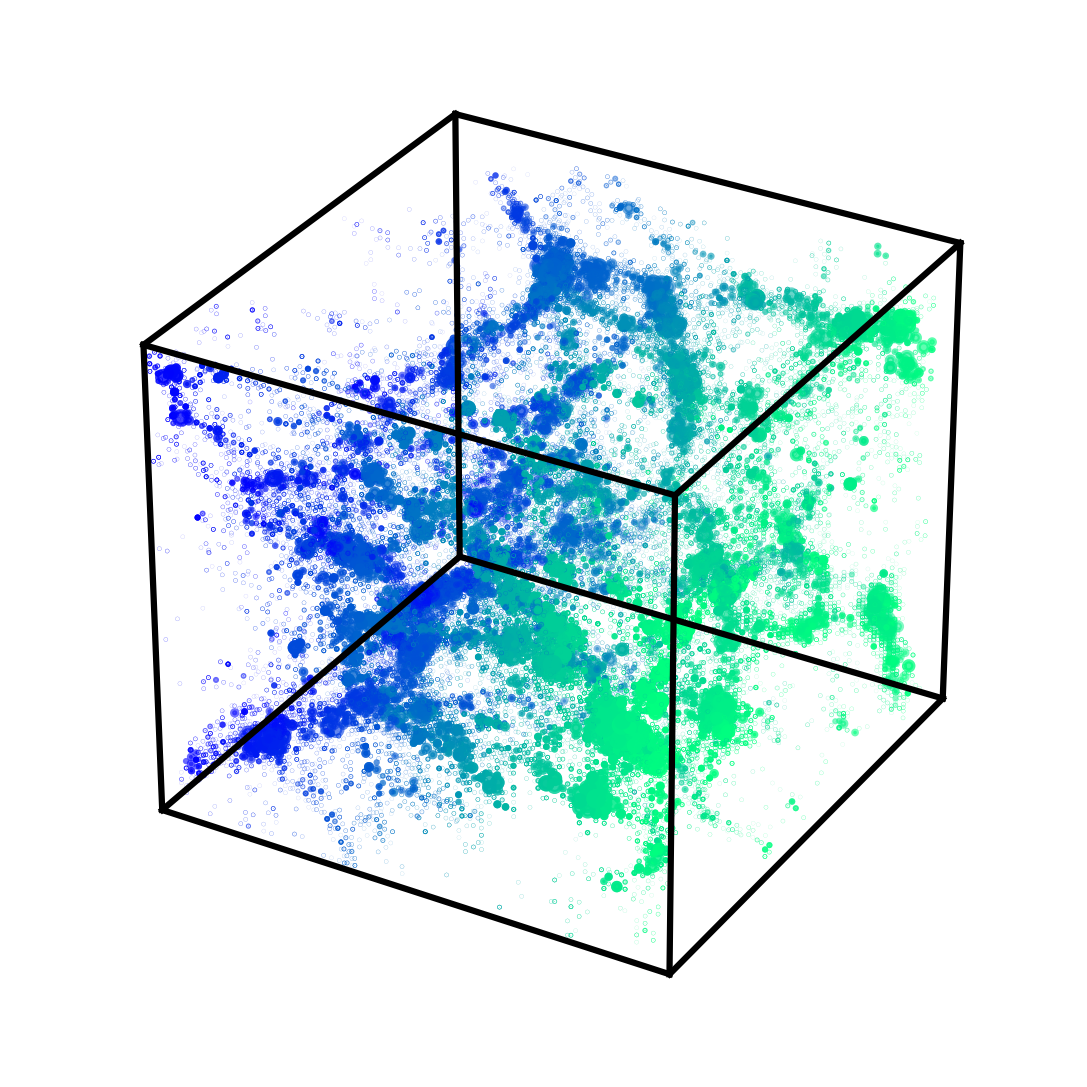}

\includegraphics[width=0.3\textwidth]{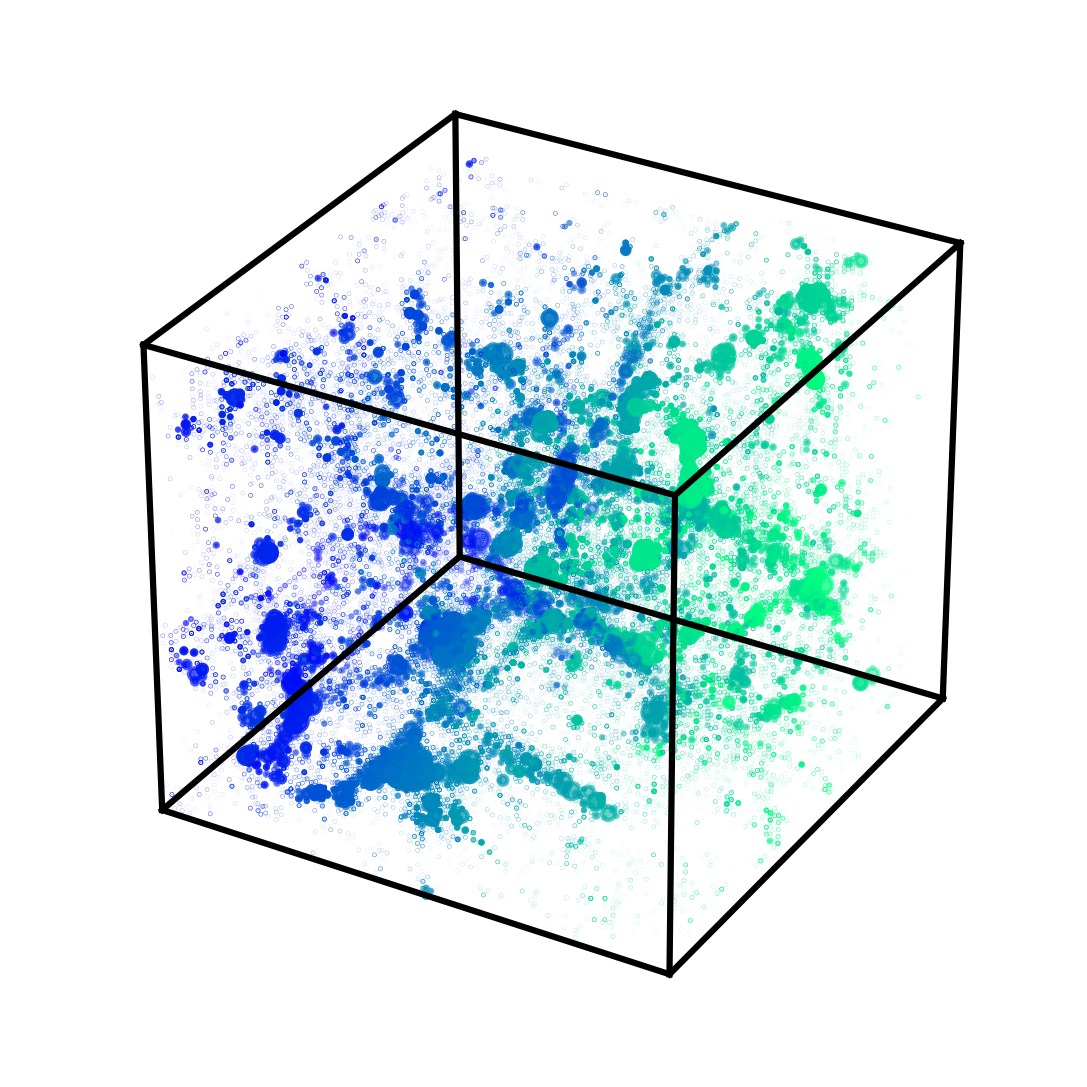}
\includegraphics[width=0.3\textwidth]{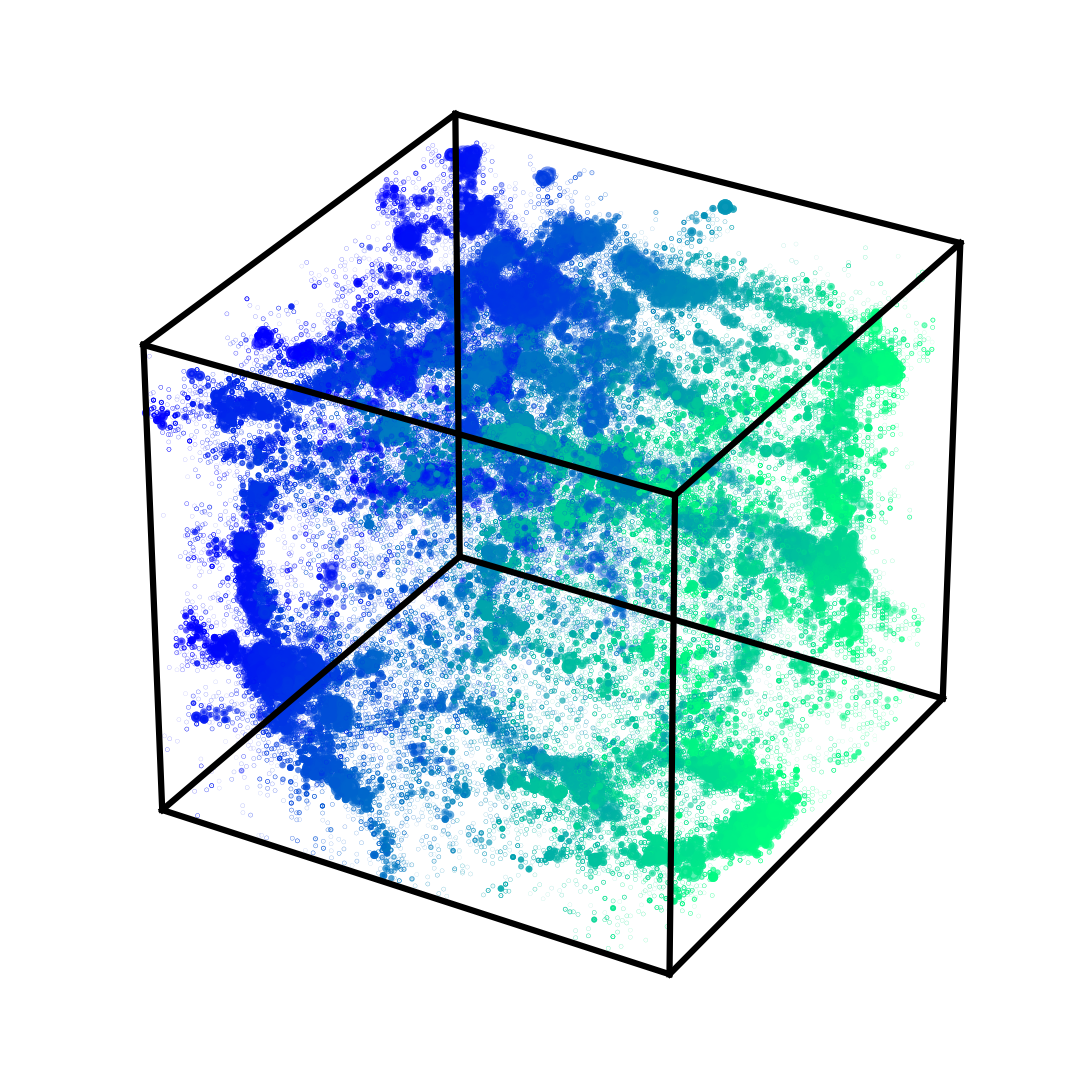}
\includegraphics[width=0.3\textwidth]{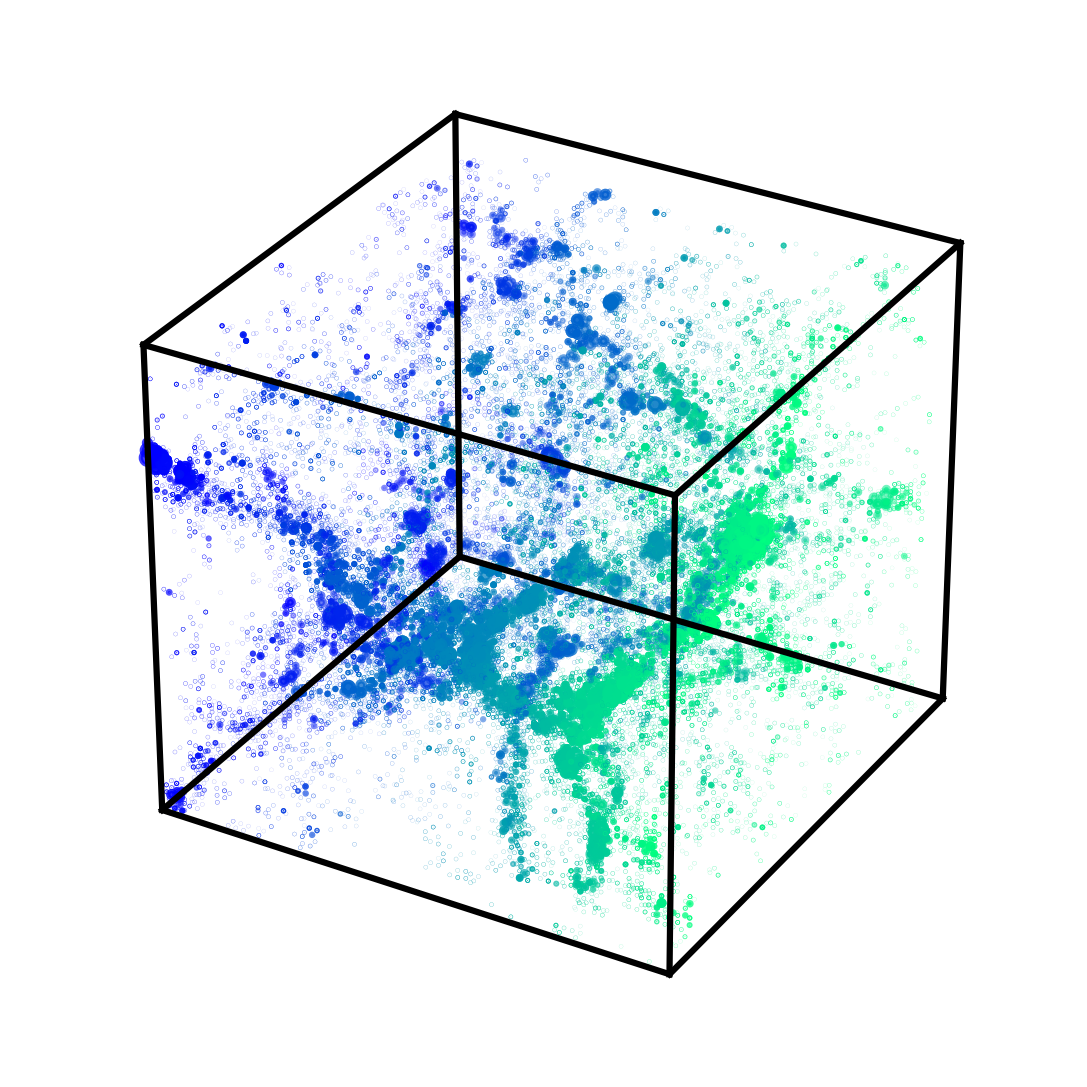}

\caption{3D visualizations of GADGET sub-volumes evolved to redshift z=0 (top row) and samples generated with a DCGAN (bottom row). Each sub-volume is visualized by plotting a number of points within the volume with size $R = (c(x)+1)^5$, where $c(x)$ is the scaled density from \eqref{scale_equation} for a given location. This is done to highlight the relevant features of large scale structure. Color scale denotes position along the $x$-axis.}
\label{fig:real_fake_samps}
\end{figure*}

\section{Results} \label{sec:results_z0}

We first present the results of our model trained on the full N-body data set evolved to $z=0$ (present day). In this section, the performance of our models is evaluated directly on the standard cosmological summary statistics. These are more physically interpretable than discriminator-based metrics such as the Frechet Inception Distance (FID) or the Inception Score (IS) that are common in the machine learning literature. For $z=0$, the model is trained on data that are scaled using \eqref{scale_equation} with parameter $\kappa=45$. In our experiments, $\kappa=45$ was the largest value where training was stable. Unless specified otherwise, our results use $N_s=2000$ samples to evaluate each GAN model and the full training set for GADGET-2 ($N_s=16384$).

Figure \ref{fig:real_fake_samps} shows visualizations of $(64 \text{ Mpc})^3$ volumes from GADGET-2 (top row) and from a trained GAN (bottom row). Samples from the generator network contain many of the filamentary structures known to populate the cosmic web. At $(64 \text{ Mpc})^3$, samples share similar types of features but also exhibit considerable variation by eye. The large variation on these scales is expected, as the homogeneity scale of the universe is typically accepted as $\gtrsim 100 \text{ h}^{-1}\text{Mpc}$ \citep{homogeneity_18, homogeneity_18_2}. Some periodic artifacts can be seen in synthetic samples, however these are low density features that do not significantly affect the cosmological summary statistics shown in the following sub-sections. 

\subsection{Voxel distribution function}
\begin{figure}
    \centering
    \includegraphics[width=\linewidth]{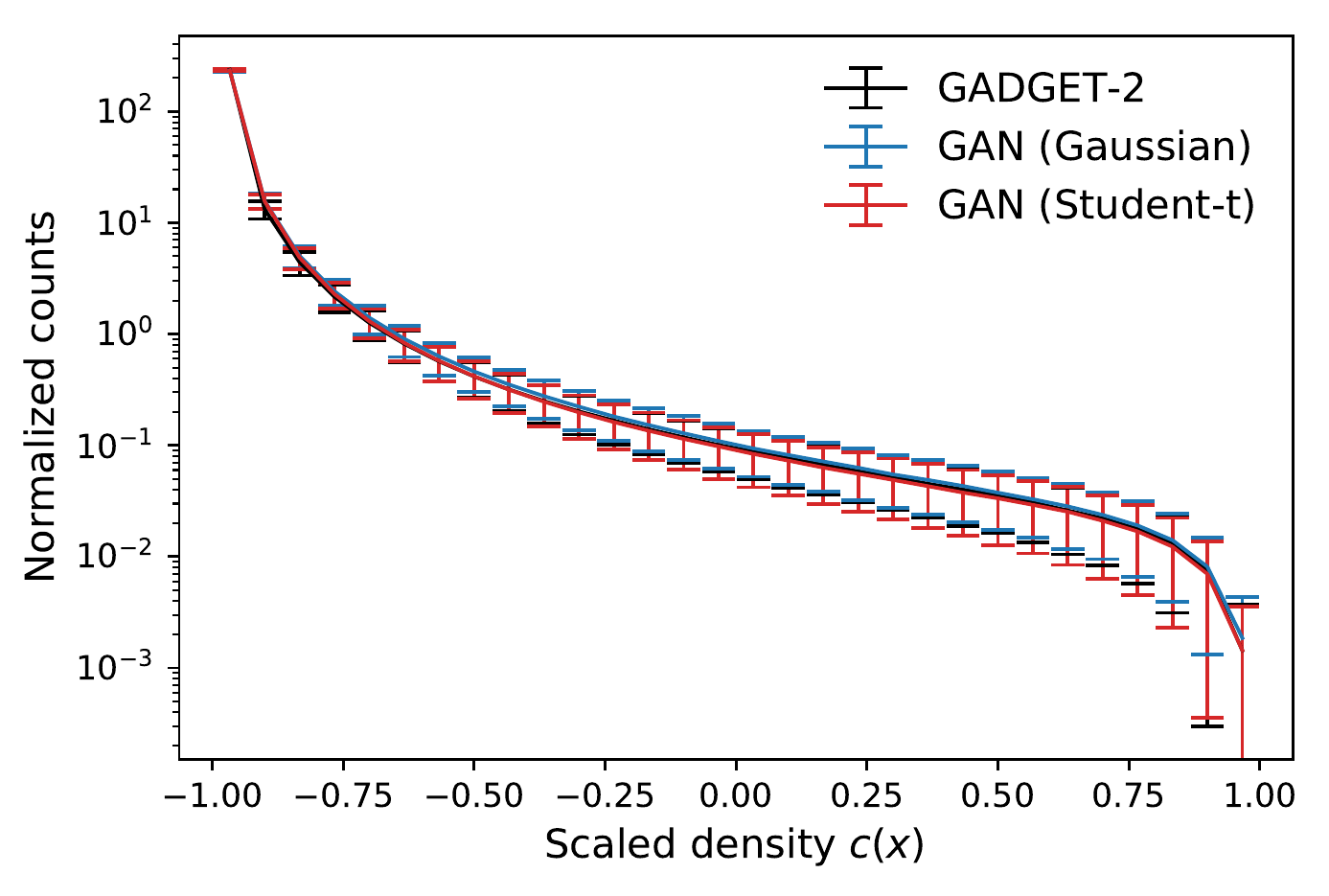}
    \includegraphics[width=\linewidth]{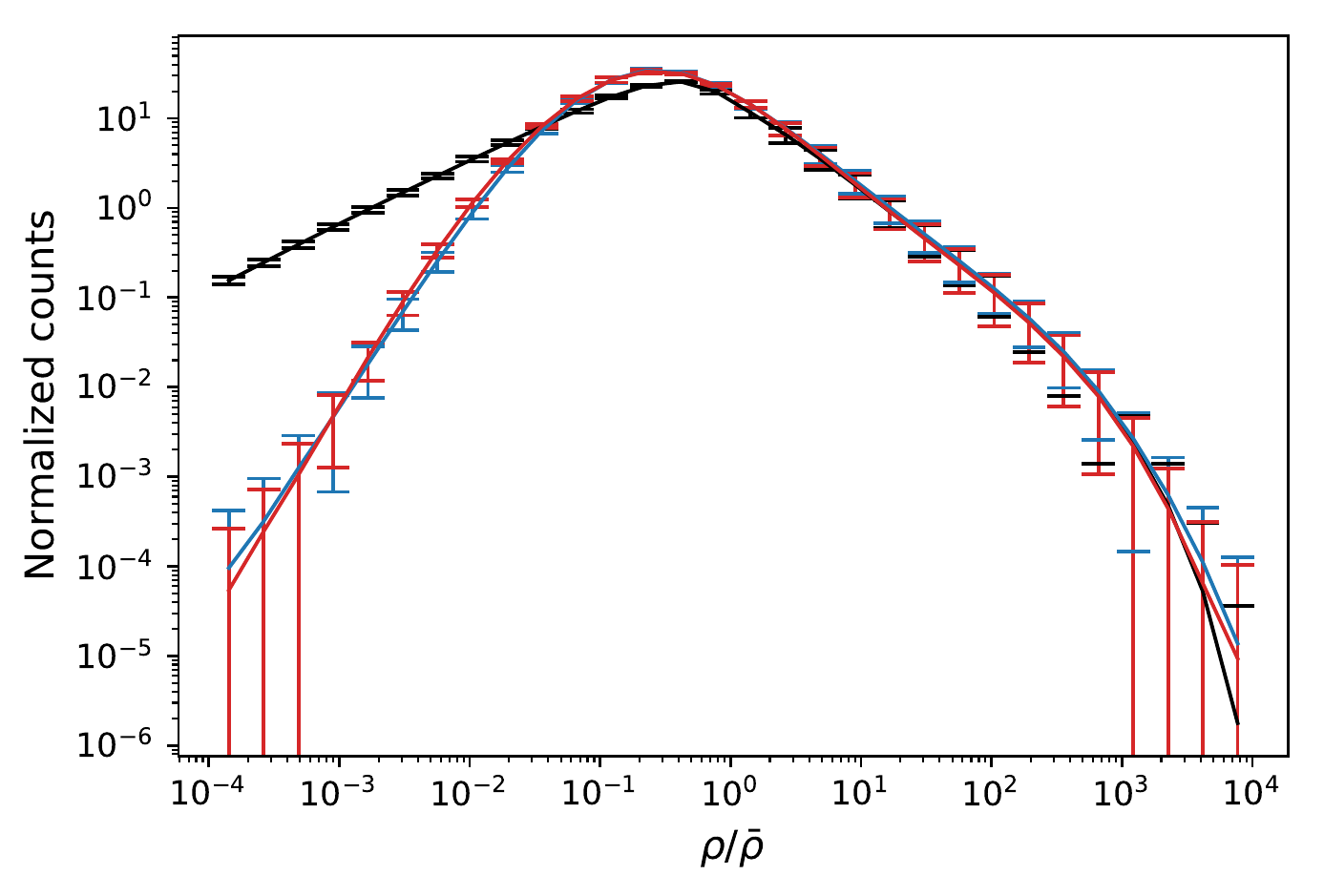}
    \caption{Voxel probability distribution functions for scaled densities $[-1, 1)$ (top) and original matter densities (bottom) of $z=0$ GADGET-2 (black) and GAN-generated sub-volumes. Blue error bars correspond to a GAN model trained with a Gaussian latent space prior, while red error bars indicate results using a Student's t-distribution prior. Each set of error bars indicates the $1\sigma$ standard deviation within each bin, computed from 2000 samples.}
    \label{fig:voxel_pdf}
\end{figure}

Figure \ref{fig:voxel_pdf} shows voxel histograms of the scaled densities (top) and normalized matter densities (bottom) for real and synthetic samples. Both types of densities are shown to highlight the fact that while the GAN trains on data scaled between $[1,-1)$, the re-scaling back to matter densities may produce unanticipated features. By using a larger scaling parameter for the data transformation, we prioritize information encoded in high-density voxels. This is reflected in the N-body and GAN-generated voxel PDFs, which agree well for the upper four orders of magnitude in density and diverge below $\rho/\bar{\rho} \sim 10^{-1}$. The deviation at low density is not evident when looking at the scaled voxel PDFs alone, which appear consistent across the full range.

The voxel PDFs in Figure \ref{fig:voxel_pdf} also highlight the difference between models that use a Gaussian latent distribution and a heavy-tailed distribution. As alluded to in the previous section, initial tests produced models that generally underestimated the variance of the training set. Furthermore, this underestimated variance was present across several summary statistics. A similar phenomenon is observed and noted in \cite{higan}. If the variance of high-density voxels is underestimated, one should expect the variance of subsequent $n$-point statistics to also be underestimated. A number of potential remedies to this problem were attempted. Adding convolutional layers to the generator/discriminator did not yield noticeable improvements, nor did modifications of the size of the latent space, the number/size of filters in the networks, the scaling of the data or the training dynamics. However, one can see that the heavy-tailed GAN more faithfully captures variance in the high-density voxel PDF. 

\begin{figure}[h]
    \centering
    \includegraphics[width=\linewidth]{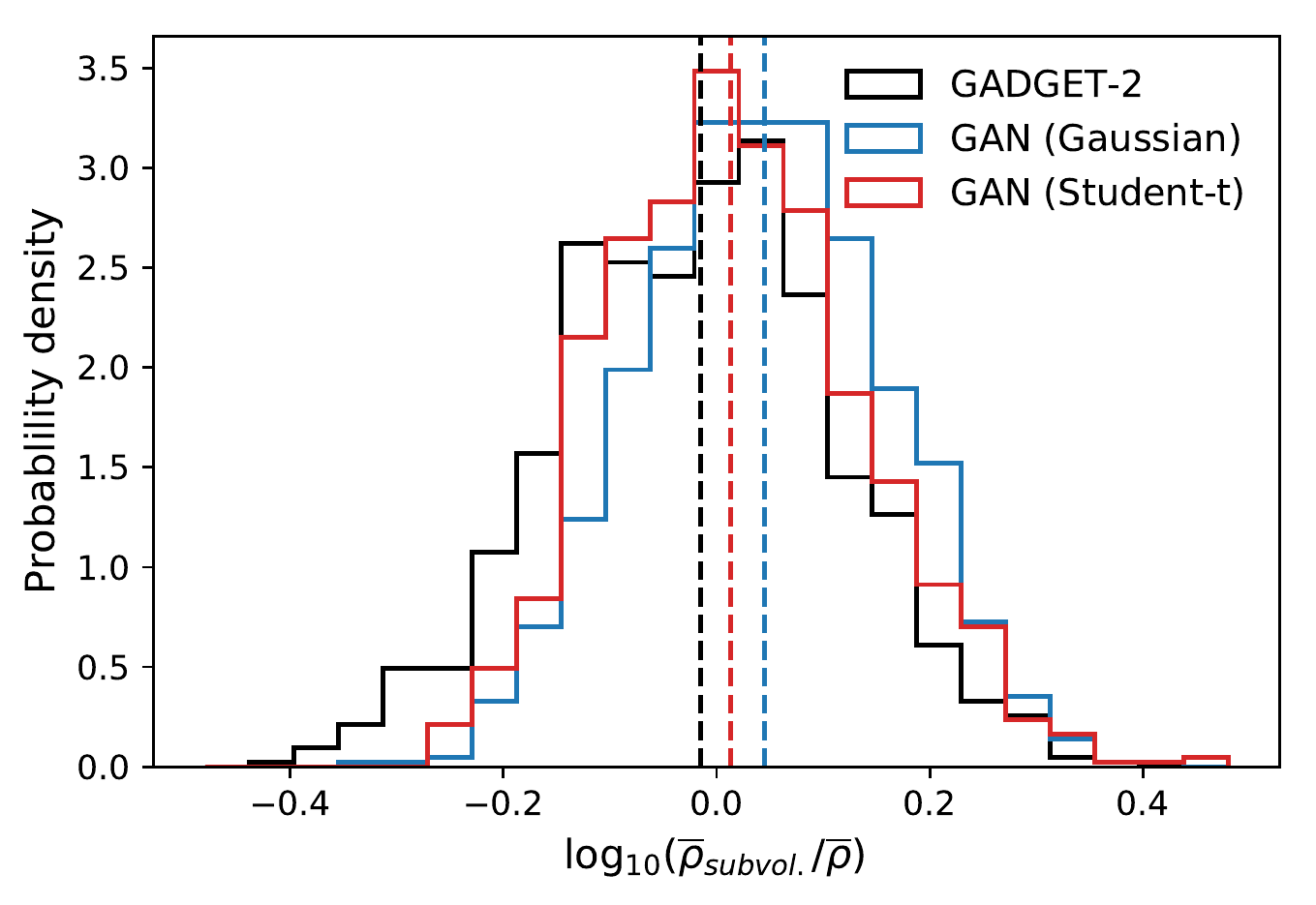}
    \caption{Log-scaled average matter densities for $z=0$ GADGET-2 sub-volumes (black) and samples generated from GANs with Gaussian (blue) and Student's t-distribution (red) latent space priors. Dashed lines indicate the median average density of each distribution. Matter density PDFs are each calculated from 2000 samples.}
    \label{fig:av_density}
\end{figure}

One validation for generative models of the matter density field is mass conservation. On the one hand, the question of mass conservation in GAN-generated simulations is ill-posed because there is no initial density field the model uses as input. However, we can examine how the distribution of mean densities $P(\overline{\rho}_{\text{subvol.}})$ for an ensemble of generated sub-volumes compares to that of N-body simulations\footnote{To clarify, $\overline{\rho}_{\text{subvol.}}$ denotes the average density of a given $(64\text{ Mpc})^3$ sub-volume, and will be different from the average density of a full GADGET-2 simulation. For the training set, each $\overline{\rho}_{\text{subvol.}}$ is computed from an already normalized matter density field $\widetilde{\rho}(x)$.}. The average density of a sample is a weighted integral over the voxel PDF, so mismatches in the real/synthetic voxel PDFs should lead to changes in the distribution of average densities. Figure \ref{fig:av_density} compares estimates of $P(\overline{\rho}_{\text{subvol.}})$ from real and synthetic samples. Samples from the GAN trained with a Gaussian latent distribution have average densities $\sim 12\%$ higher compared to samples in the training set. This is consistent with the fact that on average, the high-density voxel PDF contains more probability mass than that of the training set. The heavy-tailed GAN does slightly better in this regard, with an median average density $\sim 5\%$ higher than that of the training set.

\subsection{Power spectrum}
\begin{figure}[h]
    \centering
    \includegraphics[width=\linewidth]{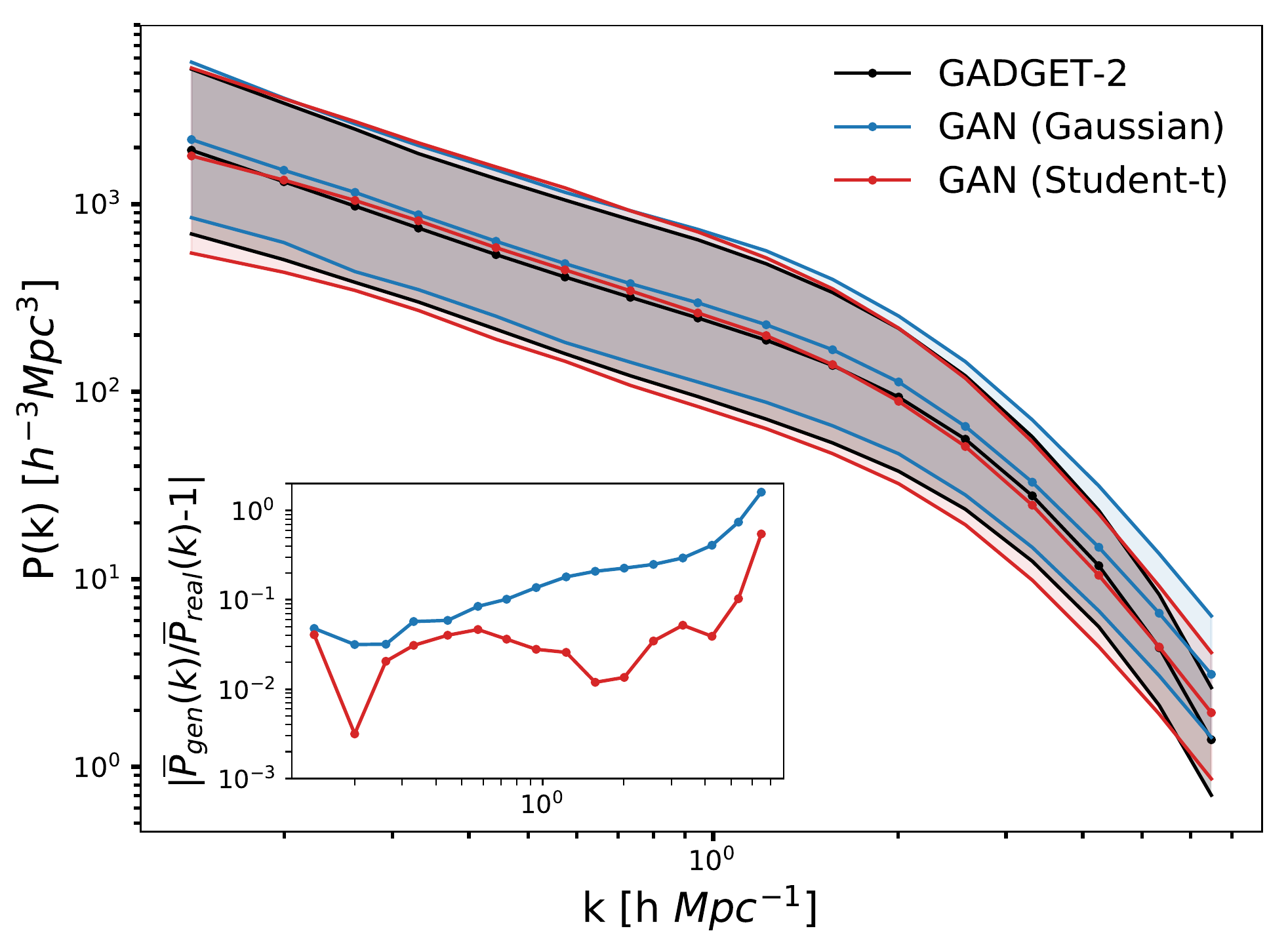}
    \caption{Comparison of (log-scaled) power spectra between GADGET-2 sub-volumes at $z=0$ (black) and samples drawn from GANs with Gaussian (blue) and Student's t-distribution (red) latent space priors. Shaded regions bound the \nth{16} and \nth{84} percentiles of the sample ensembles for each $k$ bin. These percentiles are calculated from 2000 samples by each GAN model and from the full GADGET-2 data set. Bottom left corner: fractional deviations of mean power spectra between GAN and GAGDET-2 samples.}
    \label{fig:power_spec_z0}
\end{figure}

Figure \ref{fig:power_spec_z0} shows a comparison of real and synthetic median power spectra, along with their \nth{16} and \nth{84} percentiles. The power spectrum is the Fourier transform of the two point correlation function $\xi(r)$: 
\begin{equation}
    P(|\vec{k}|) = \int \xi(\vec{r})e^{-i \vec{k}\cdot \vec{r}}d^3 \vec{r}.
\end{equation}
where $\xi(|\vec{r}|) = \langle \delta(\vec{r}')\delta(\vec{r}'+\vec{r})\rangle$, $k = 2\pi/\lambda$ is the fluctuation wavenumber and $\delta(\vec{r})=\rho(\vec{r})/\overline{\rho} - 1$ is the density field contrast. We use the software package \texttt{nbodykit} \citep{nbodykit}, which uses a Fast Fourier Transform (FFT) method to compute the power spectra of our samples. Also plotted are the mean absolute fractional deviations of synthetic power spectra from GADGET-2 power spectra, i.e. $\left|(\overline{P}_{gen}(k)/\overline{P}_{real}(k)) - 1\right|$.

At the physical scales considered in this work, the CDM power spectrum is highly non-linear. In particular, $P(k)_{nonlin}/P(k)_{lin}$ grows from unity at $k\sim 0.1$ to over a factor of ten for $k \geq 1$ \cite{dark_universe}. Both GANs produce samples that are accurate well into the non-linear regime. However, the heavy-tailed GAN produces more accurate power spectra, with significant improvement at large $k$. In particular, the mean power spectra from GADGET-2 and our heavy-tailed GAN agree within $5\%$ for $k \leq 3$ and to $<10\%$ up to $k=5$. At these scales, baryonic effects have a significant impact on the matter power spectrum, and so we do not claim to capture the full matter power spectrum here. Nonetheless, our results on dark matter only simulations demonstrate the ability of our generative models to capture highly non-linear behavior on small scales. The ensemble of power spectra also captures super-sample variance from modes larger than 64 Mpc, since our training samples are sub-volumes from larger simulations.
\begin{figure}[h]
    \centering
    \includegraphics[width=\linewidth]{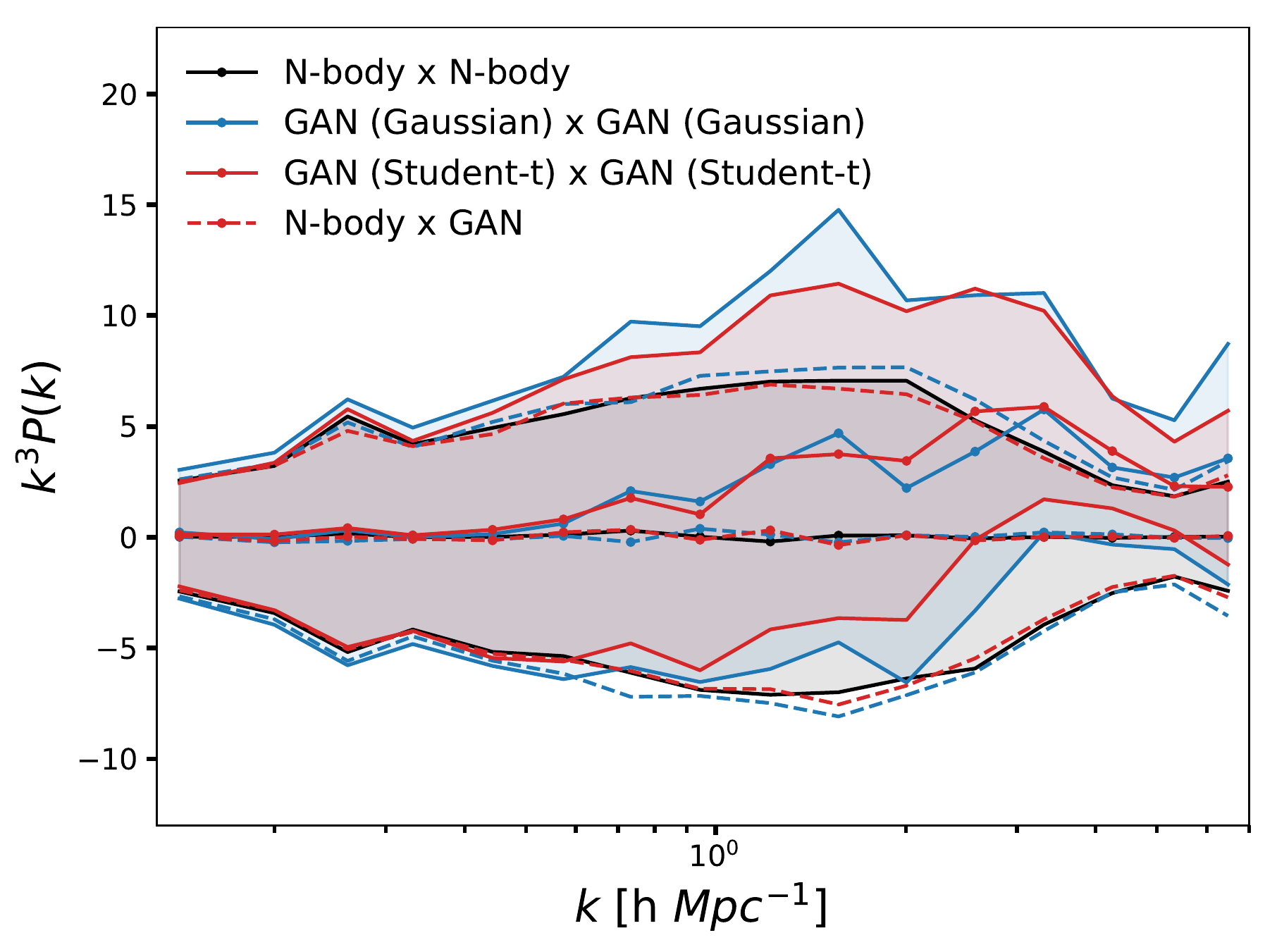}
    \caption{Cross-power spectra for various combinations of GADGET-2 (black) and synthetic (blue, red) N-body sub-volumes. Dashed lines indicate cross power spectra between GAN generated samples and GADGET-2 sub-volumes. Each cross-spectrum combination is evaluated using 2000 sub-volume pairs. Shaded regions bound \nth{16} and \nth{84} percentiles of for each sample ensemble and within each $k$ bin.}
    \label{fig:auto_cross_spec}
\end{figure}
\subsubsection{Covariance and Correlation Matrices}
\begin{figure*}[t]
\centering
\includegraphics[width=0.87\textwidth]{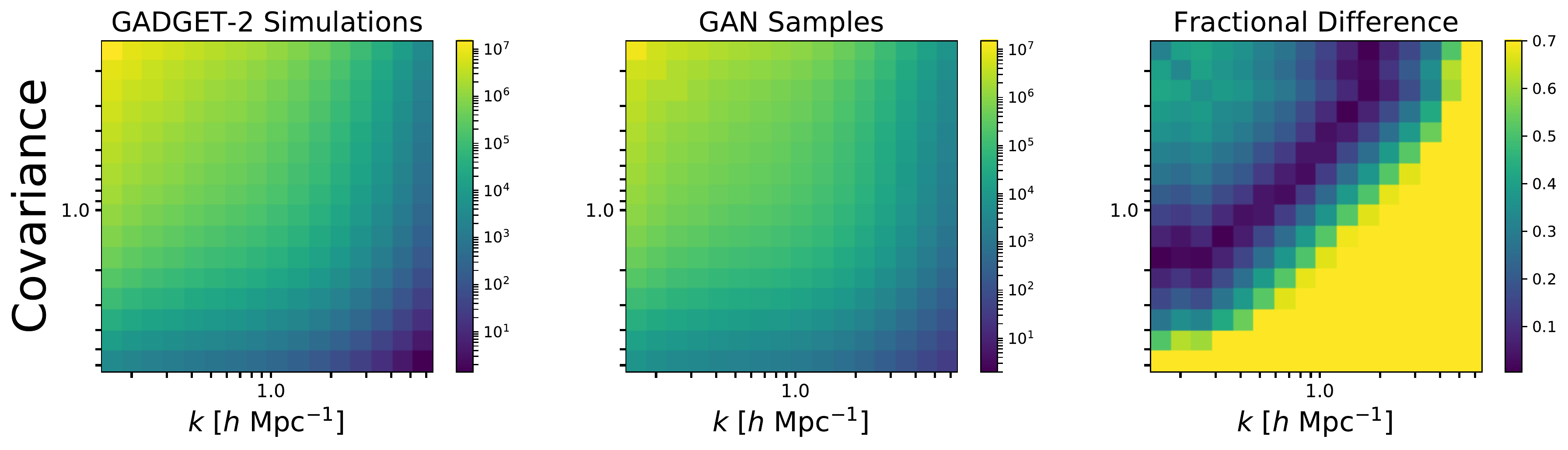} 
\includegraphics[width=0.87\textwidth]{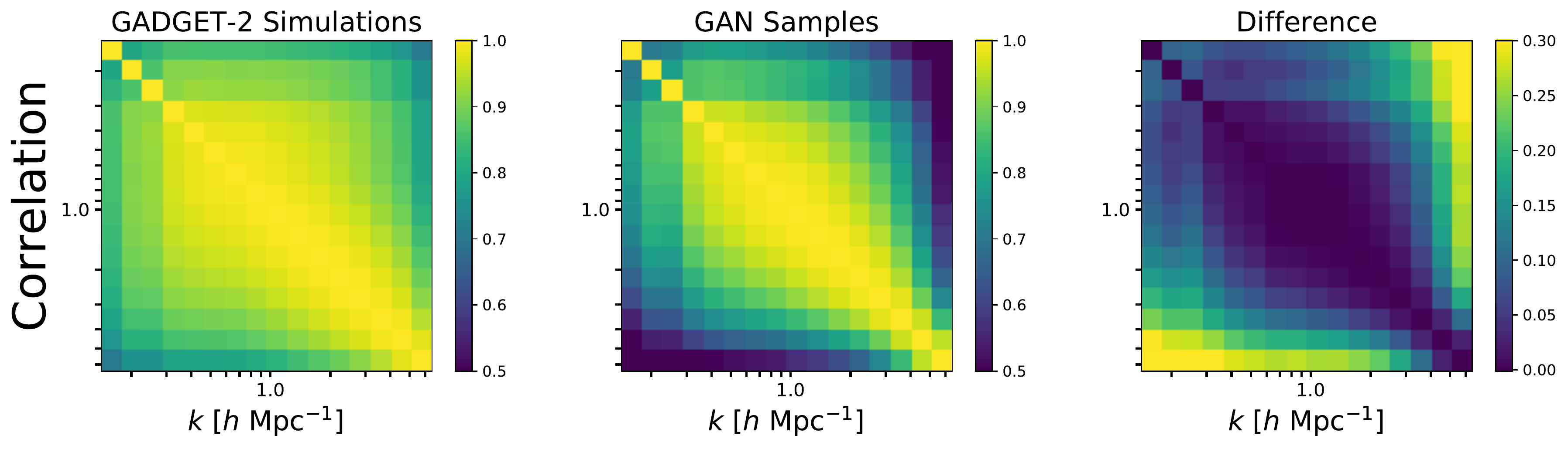}
\caption{\textbf{Top row:} Power spectrum covariance matrices of GADGET-2 N-body sub-volumes (left), synthetic samples drawn from a GAN with Gaussian latent space prior (middle), and their fractional difference (right). \textbf{Bottom row:} same as top row but for power spectrum correlation matrices and their difference. The covariance and correlation matrices from our GAN model are computed from 2000 synthetic samples, while those of GADGET-2 are computed using the full training set.}

\label{fig:cov_corr_gaussian}
\centering
\includegraphics[width=0.87\textwidth]{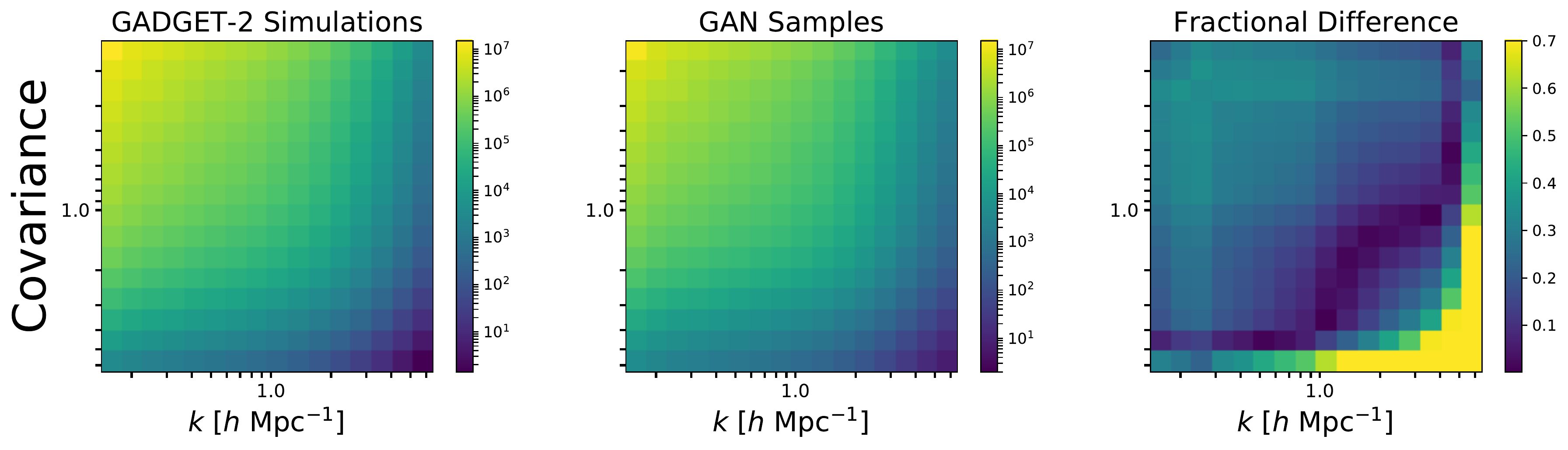} 
\includegraphics[width=0.87\textwidth]{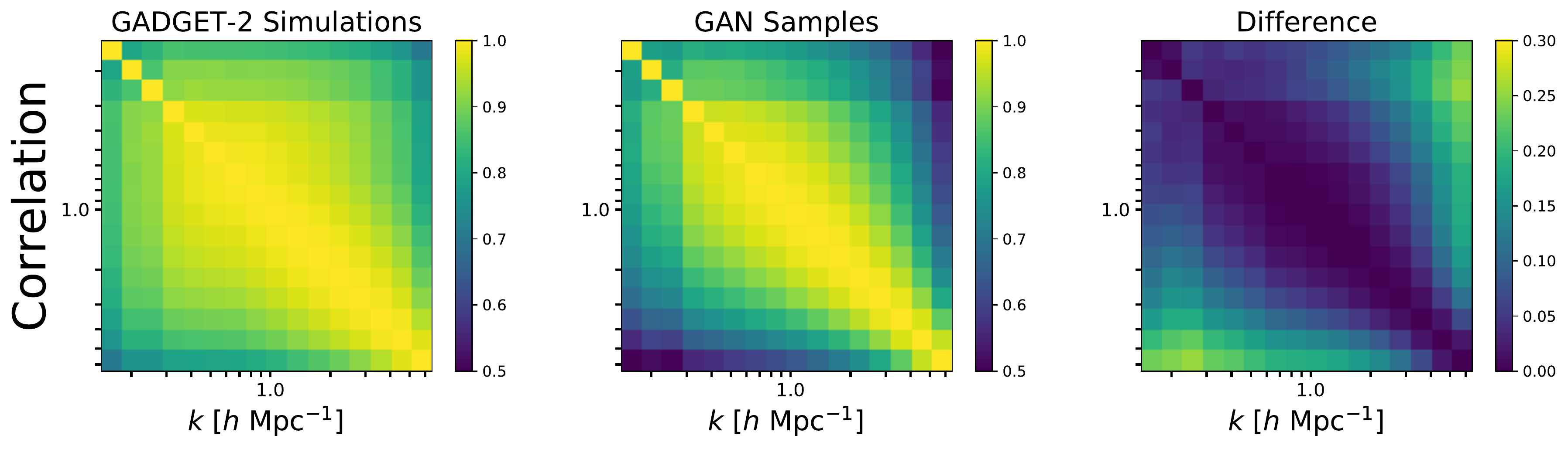}

\caption{The same as Figure \ref{fig:cov_corr_gaussian}, but for a GAN where the latent space prior is modeled with a Student's t-distribution.}

\label{fig:cov_corr_student_t}
\end{figure*}

Figures \ref{fig:cov_corr_gaussian} and \ref{fig:cov_corr_student_t} show the covariance and correlation matrices for samples generated by Gaussian and heavy-tailed GANs, respectively, along with their deviation from GADGET-2 N-body samples. We compute power spectrum covariance matrices with the sample covariance estimator
\begin{equation}
    \hat{\mathcal{C}}_{ij} = \frac{1}{N_s-1}\sum_{n=1}^{N_s}\left[P_n(k_i)-\overline{P}(k_i)\right]\left[P_n(k_j)-\overline{P}(k_j)\right],
\end{equation}
in which $N_s$ is the number of samples, $P_n(k)$ is the dark matter power spectrum of sample $n$ and $\overline{P}(k)$ is the mean power spectrum across samples. The correlation matrix is further defined as
\begin{equation}
    \hat{R}_{ij} = \frac{\hat{C}_{ij}}{\sqrt{\hat{C}_{ii}\hat{C}_{jj}}}.
\end{equation}
The improvement from using a heavy-tailed prior is especially apparent when comparing covariance matrices. The covariance matrices computed from the heavy-tailed GAN are over $50\%$ more accurate on all scales, agreeing to between $5\%-25\%$ up to $k=3$. The heavy-tailed GAN also improves the estimated correlation matrices on all scales, with notable improvement for correlations between large and small scales. At high $k$, covariance matrices from both GANs overestimate the true covariance, which is consistent with overestimates of the small-scale power spectrum seen in Figure \ref{fig:power_spec_z0}. Both sets of synthetic correlation matrices follow the structure of the N-body correlation matrices. This is the first work to present power spectrum covariance matrices of GAN-generated 3D dark matter density fields. 

\subsubsection{Cross-power spectrum}
Cross-power spectra help quantify the degree to which generated samples are correlated with one another and with the training set at different spatial scales. Figure \ref{fig:auto_cross_spec} shows N-body $\times$ N-body, N-body $\times$ GAN and GAN $\times$ GAN cross spectra, each evaluated from 2000 pairs of samples. Setting aside any large-scale effects that might come from evaluating sub-volumes of the same N-body simulation, a random N-body sample should not (statistically) have any aligned structure with another random sample, i.e. the average cross spectrum should be zero. While the cross spectra between real and synthetic samples are consistent with zero, the non-zero GAN $\times$ GAN cross-spectra observed for $k \geq 0.5$ indicate that our models produce some coherent structure across samples on these scales. 

\subsection{Bispectrum}

\begin{figure*}
\centering
\includegraphics[width=0.4\textwidth]{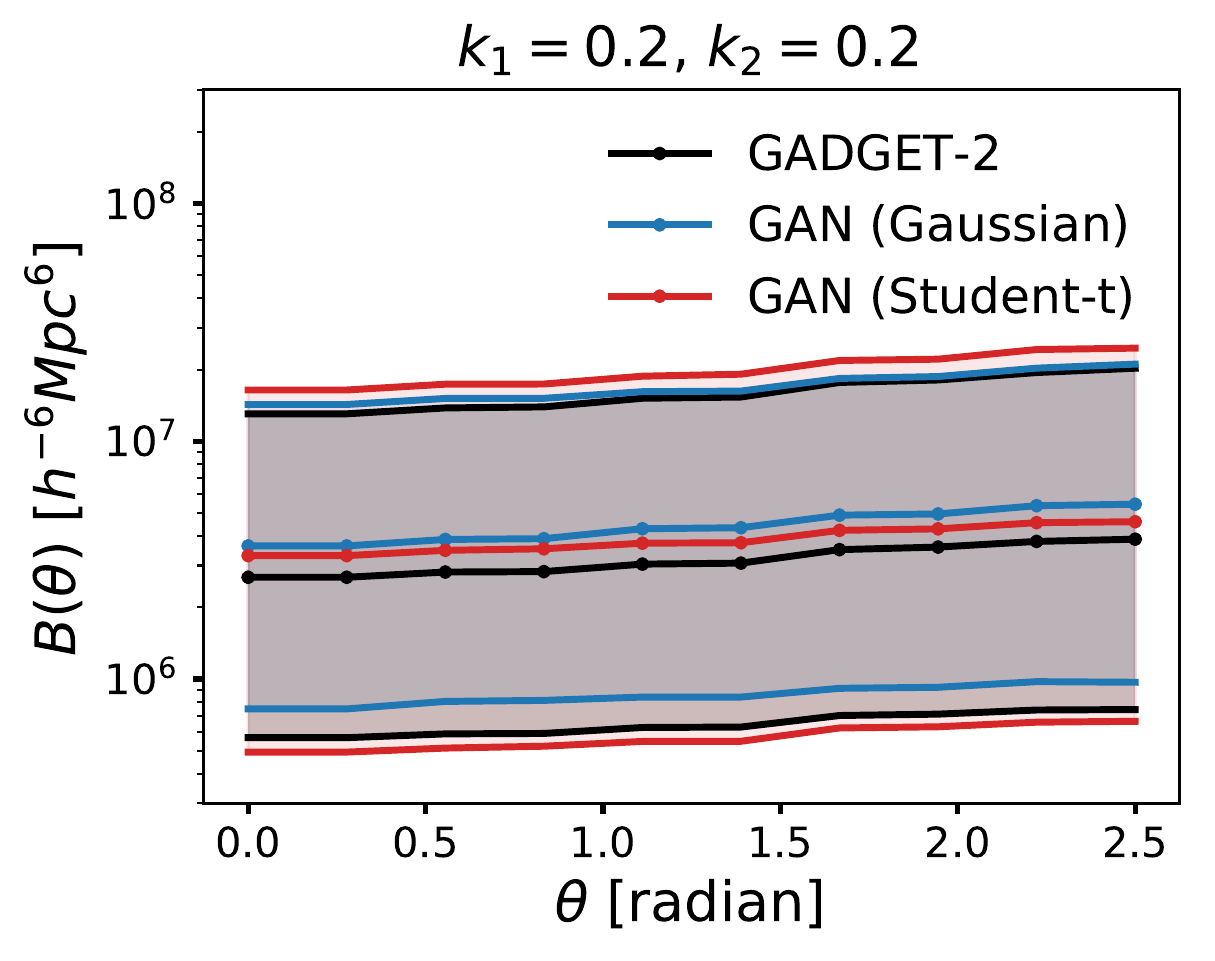}
\includegraphics[width=0.4\textwidth]{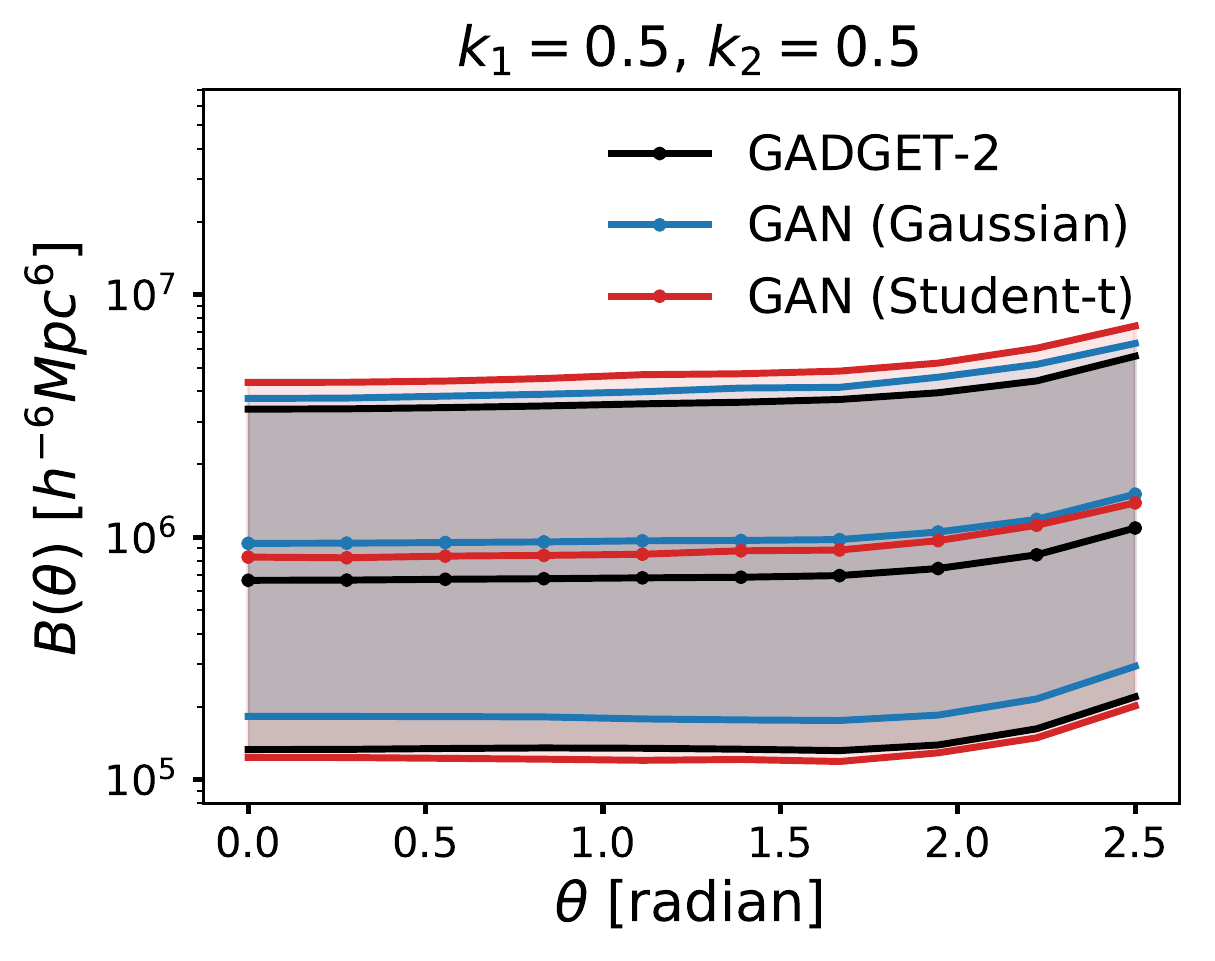}
\includegraphics[width=0.4\textwidth]{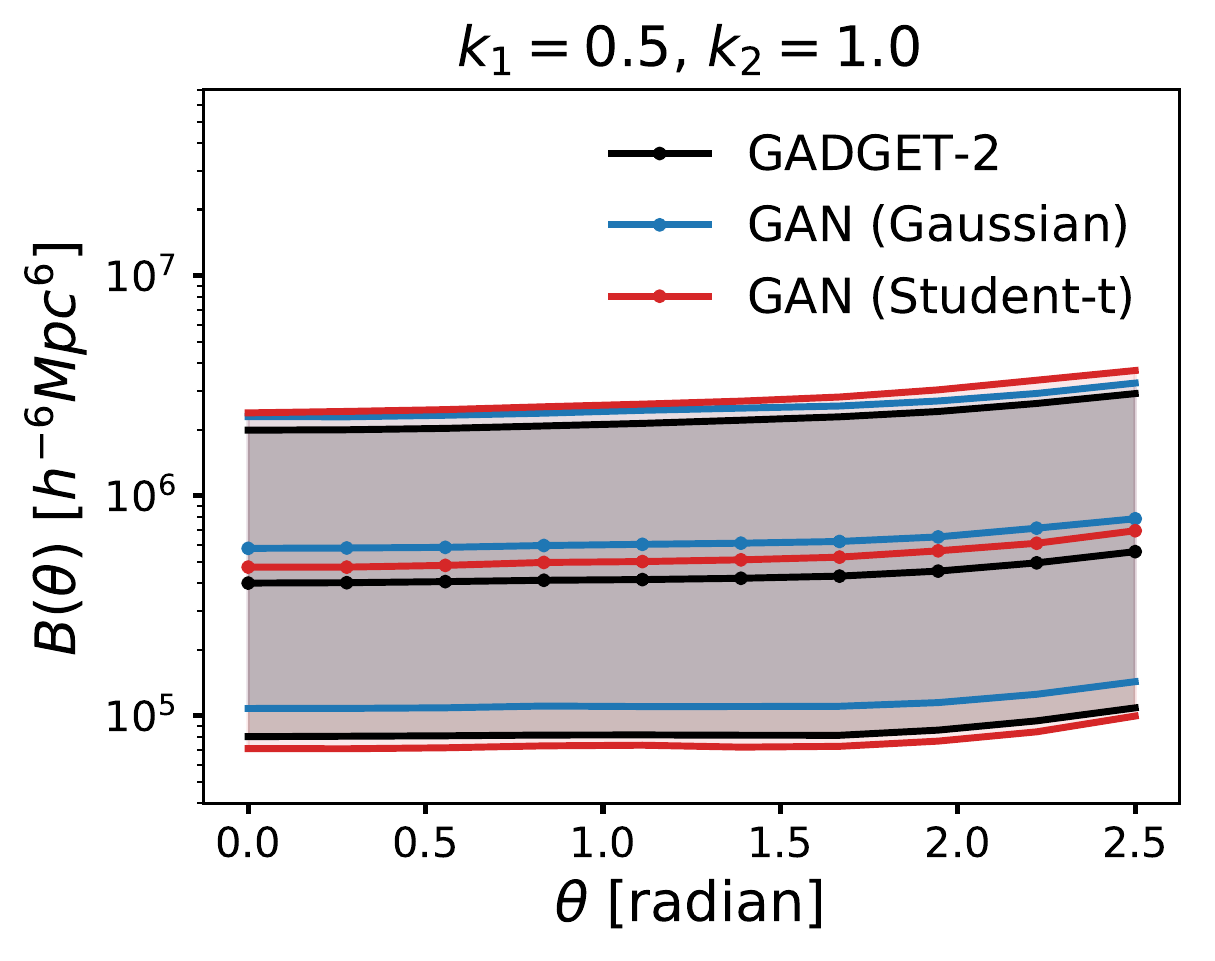}
\includegraphics[width=0.4\textwidth]{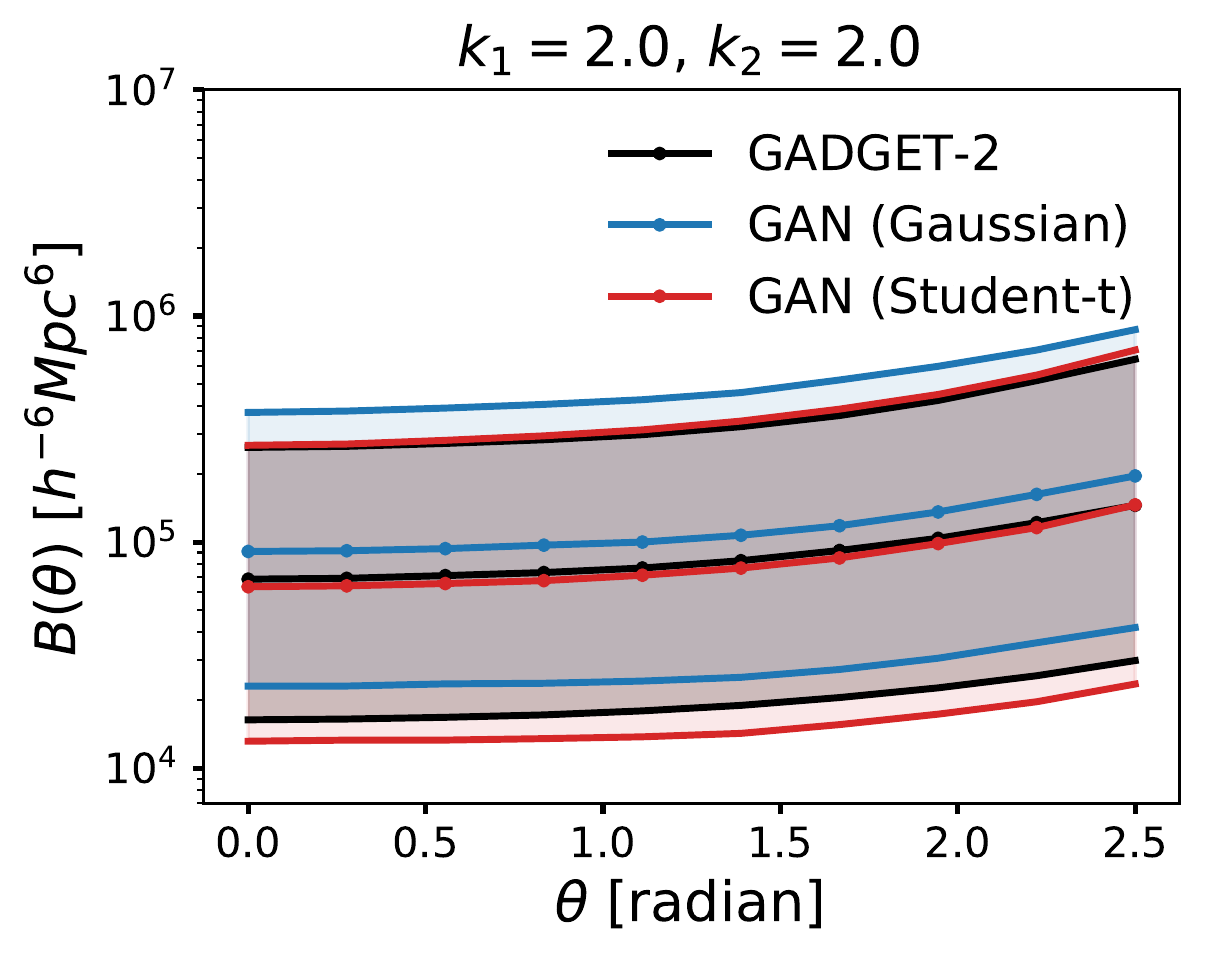}
\caption{Comparison of real (black) and synthetic (blue, red) matter bispectra for isosceles and other triangle configurations. Central lines indicate ensemble medians, while shaded regions bound the \nth{16} and \nth{84} percentiles for each set of bispectra. These percentiles are calculated from 2000 samples for each GAN model and from the full GADGET-2 data set.}
\label{fig:bispectra}
\end{figure*}

A comparison of matter bispectra allows us to verify that our trained GANs learn to accurately model non-linear features of the cosmic web. The bispectrum is defined as the Fourier transform of the three-point correlation function:
\begin{equation}
(2\pi)^3 B(\vec{k}_1,\vec{k}_2,\vec{k}_3)\delta_D(\vec{k}_1+\vec{k}_2+\vec{k}_3) = \langle \delta(\vec{k}_1)\delta(\vec{k}_2)\delta(\vec{k}_3)\rangle
\end{equation}
where $\delta_D$ is the Dirac delta. Figure \ref{fig:bispectra} shows bispectra calculated from real and synthetic N-body samples, for a variety of large- and small-scale triangle configurations. We calculate sample bispectra using the software package \texttt{Pylians}\footnote{\url{https://github.com/franciscovillaescusa/Pylians}}. The accuracy of the median generated bispectra is best at small scales and gets slightly worse for larger triangle configurations (in physical space), though notably these errors appear constant with respect to $\theta$, the angle between wave-vectors. The angular dependence of each bispectrum configuration also appears to be captured by the \nth{16} and \nth{84} percentiles of the GAN sample ensembles. Overall, the heavy-tailed GAN yields a more accurate bispectrum distribution across samples, consistent with our earlier comparisons of matter power spectra and voxel density distributions.

\subsection{Interpolation over redshift}

In addition to single redshift sample generation, we consider GAN models that interpolate in a latent space augmented by conditional redshift information. In particular, we seek to establish that GANs can smoothly interpolate in redshift space for a given volume. For this task, we train on snapshots of twenty-four $(512\text{ Mpc})^3$ GADGET-2 volumes at redshifts $z=3.0, 1.5, 0.5$ and 0.0, which gives us $\sim 48000$ training samples. While the scaling parameter $\kappa=45$ worked well for N-body simulations at $z=0$, such a scaling does not effectively capture information at higher redshifts, when the matter density field is closer to Gaussian. Indeed, attempts to train our conditional redshift model with $\kappa=45$ were unsuccessful and often ended in divergent training dynamics. We instead use $\kappa=4$, which allows for stable training, albeit at the cost of accuracy on small scales. 

\begin{figure*}[t]
    \centering
    \includegraphics[width=0.95\textwidth]{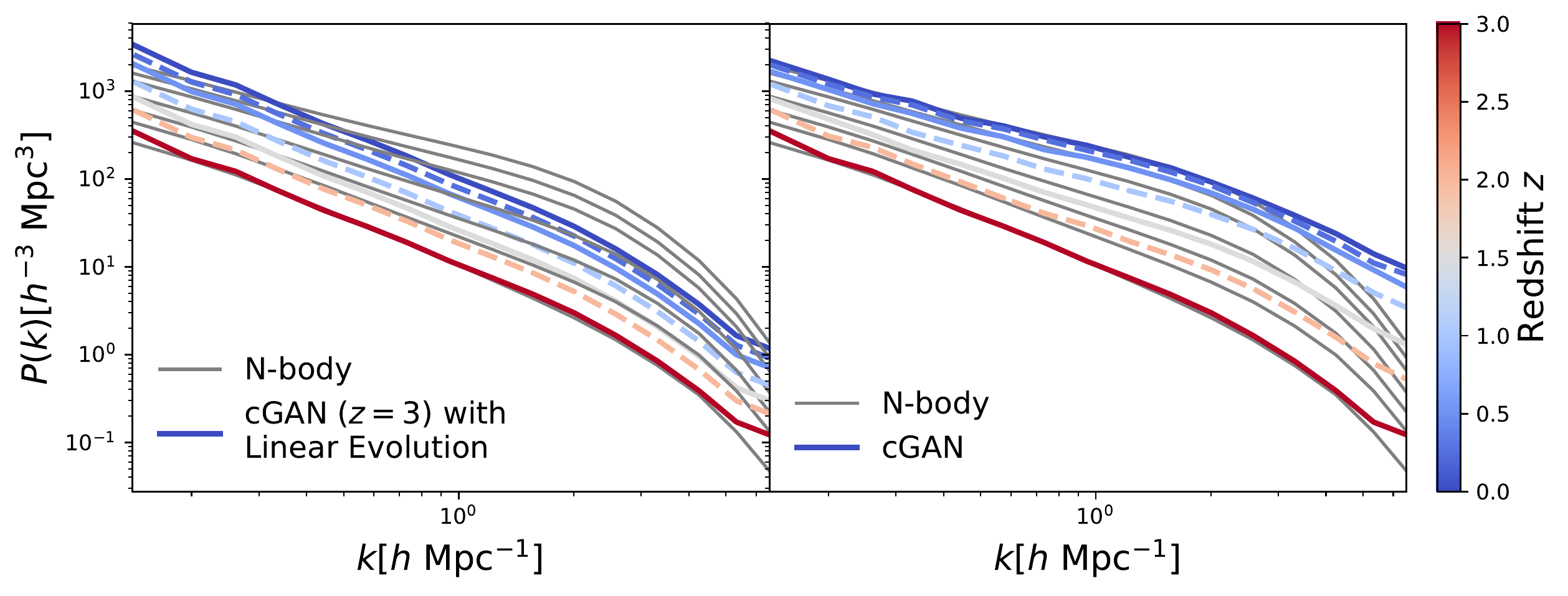}
    \caption{\textbf{Left:} N-body power spectra (gray lines) compared with a median initial cGAN power spectrum at $z=3$ that is evolved with linear theory (colored lines). \textbf{Right:} the same N-body power spectra, compared with cGAN-interpolated median power spectra. Power spectra in both panels are evaluated at both training redshifts ($z=3.0, 1.5, 0.5, 0.0$, solid lines) and interpolated redshifts ($z=2.0, 1.0, 0.25$, dashed lines), with color bar denoting the conditional redshift input to the cGAN.}
    \label{fig:redshift_evol}
\end{figure*}

Figure \ref{fig:redshift_evol} shows the extent to which our conditional GAN learns a non-linear mapping between redshift and the output density field. In the left panel, an initial power spectrum generated by the conditional GAN at $z=3$ is evolved linearly,
\begin{equation}
    P(k, z) = \left( \frac{D(z)}{D(z=3)} \right)^2 P(k, z=3)
\end{equation}
where $D(z)$ is the linear growth factor, while in the right panel the cGAN-interpolated matter power spectra are shown. We evaluate our model at both the aforementioned training redshifts and at intermediate redshifts $z=2.0, 1.0,$ and 0.25. It is clear that while linear theory underestimates the growth of structure on small scales, the conditional GAN successfully captures the non-linear gravitational evolution. There is no obvious difference in cGAN performance evaluated at trained vs. interpolated redshifts, though interpolated power spectra at low redshift ($z < 1$) appear to be clustered around $z=0.25$.

\begin{figure*}
    \centering
    \includegraphics[width=\linewidth]{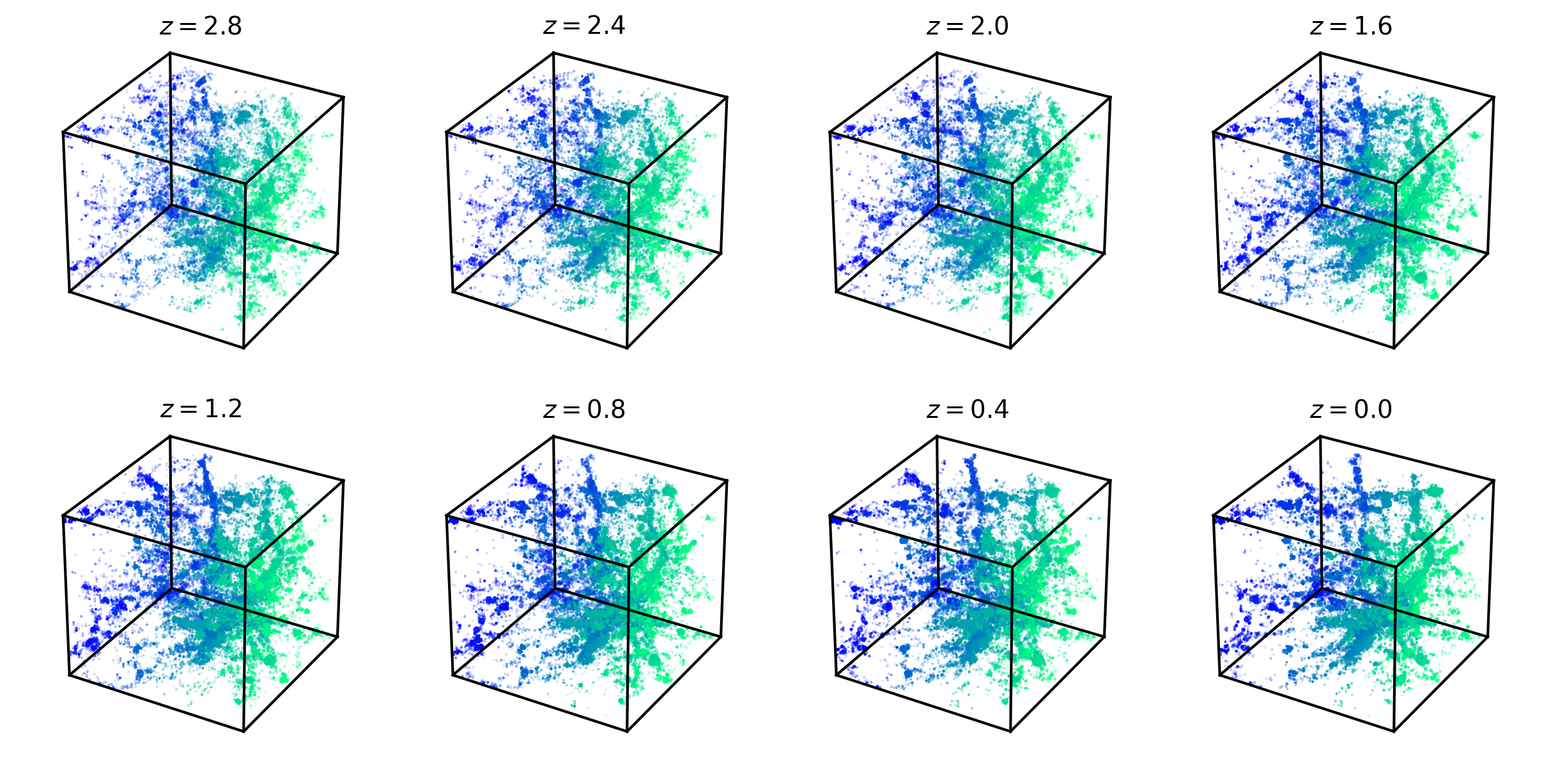}
    \caption{3D visualizations of cGAN-generated sub-volumes, with conditional redshift $z$ linearly interpolated from $z=2.8$ to $z=0$. Sub-volumes are visualized using the same scheme as in Figure \ref{fig:real_fake_samps}.}
\label{fig:diffz_samps}
\end{figure*}

Figure \ref{fig:diffz_samps} shows a sequence of cGAN-generated samples where the latent vector $\mathbf{z}$ is fixed and the conditional redshift parameter is varied from $z=2.8$ to the present day. Because we do not present consecutive redshift snapshots of the same volume to the discriminator and because we do not enforce mass conservation when varying the latent redshift parameter, we do not expect the cGAN to learn to simulate gravitational evolution per se, which would involve accurate modeling of the matter density displacement field. Nonetheless, our model learns a smooth mapping between redshift and the matter density field in real space -- regions that are mildly clustered at high redshift appear to form coherent filaments as the conditional redshift is varied, with no obvious intermediate discontinuities. This visual inspection helps establish the property that smooth paths in the latent space correspond to smooth paths in the data space, which reinforces the robustness of the compressed GAN representation.

\section{Discussion and Future Work}

In this work we have demonstrated that generative adversarial networks can learn a high-fidelity, compressed representation of the 3D dark matter density field. Once the generative model is trained, new samples can be generated in a fraction of a second -- including the overhead from memory transfer between GPU and CPU, generating 512 samples (the volume equivalent of one N-body simulation) takes $\sim 6$ seconds. Furthermore, our trained generator only takes 17 Mb of storage, and storage of individual realizations amounts to saving latent vectors that can be passed through the generator when needed. We exploit the rotational invariance of our cosmological N-body simulations by using random rotations and flips to augment the training data set, which contributes to the stability of training. In fact, we are able to stably train our GANs with as few as 2048 samples (see Appendix \ref{app:training_data}), equivalent to four $(512 \text{ Mpc})^3$ GADGET-2 simulations. 

Synthetic samples were validated using one-, two- and three-point statistics that are accurate well into the non-linear regime where baryonic effects become important. This work presents the first estimates of power spectrum covariance matrices and bispectra for GANs sampling from a compressed representation of the data. While Figure \ref{fig:auto_cross_spec} indicates some coherent structure in synthetic samples on small scales, our models largely reproduce the power spectrum covariance structure of the matter density field when compared with N-body simulations. 

Given the dynamic range of voxel densities recovered from N-body simulations, recovering accurate realizations over the full, unscaled voxel PDF is difficult. High density voxels have the largest impact on summary statistics like power spectra and bispectra, and so in this work we chose a data scaling that maximizes contrast in that regime. Some cosmological analyses may warrant more emphasis on low-density features (e.g. studies of the Lyman-alpha forest), in which case the data scaling can be modified to preserve information at those densities. However, it is unclear whether an arbitrary voxel PDF with large dynamic range can be learned, given the precision of a single network and the unsupervised nature of GAN training. One might train several networks on different scaled versions of the same data, however this is beyond the scope of this work.

This work also demonstrates that conditional GANs are capable of learning a smooth mapping between early and late times in the evolution of the matter density field, with fairly accurate interpolation as seen through the dark matter power spectrum. Future implementations might use a redshift-dependent data scaling that preserves the information in linear fluctuations at high redshift and non-linear fluctuations at low redshift, though our attempts to do so were unsuccessful. 

We also show that using a Student's t-distribution prior for the latent space results in a model that better captures sample variance across several summary statistics. While it is true that deep neural networks have high capacity, a given modeling task may be harder in the presence of distributional mismatch between the latent space and the data space. A similar point is argued in \cite{dual_ml}, which considers the role of dual data representations for classification. In that work, even when networks were given sufficient capacity to learn a dual transformation from position space to momentum space, where classification was straightforward, no trained model learned one. To our knowledge, this is the first application of heavy-tailed priors for modeling the latent space of a GAN. While we choose the Student's t-distribution as our heavy-tailed prior, we remain agnostic to the ``optimal" prior. Depending on the data set and purpose of the generative model, it may be possible to optimize the prior distribution non-parametrically as in \cite{nonparametric_gan}. We expect this modification of the standard GAN prior to be helpful for data sets with tails that impact the overall quality and diversity of the data distribution. While heavy-tailed priors are not necessarily the only remedy for issues addressed in this work, it was the only modification among several to the fiducial model and training procedure that led to models that were accurate and variance preserving. Furthermore, the change is very simple to implement -- in our work, only one line of code needed to be changed. 

Because our GANs learn a compressed representation of the data distribution, it may be possible to further de-correlate synthetic N-body ensembles by resampling the latent space of the generator after training. When one trains a GAN to model data with a lower-dimensional latent space, the underlying assumption is that a compressed representation of the data distribution exists on a manifold with dimension equal to that of the latent space. Mapping training samples to the posterior $P(\mathbf{z}|x_{data})$ is straightforward in models like variational autoencoders where the encoder network learns the mapping during training -- for GANs this may be possible through a back propagation optimization scheme like in \cite{backprop_gan}. An estimate of the training set posterior could help determine which regions of the latent space are correlated. With an approximate posterior available, one might then train a normalizing flow to resample the latent space in a decorrelated manner, as has been done with variational autoencoders in \cite{normalizing_flow_vae}. 

A promising avenue in machine learning literature seeks to embed inductive biases into neural network models, such that they are guaranteed to respect the structures and symmetries of a system \citep{spherical_cnn, gauge_cnn, homeomorphic_vae}. Such properties make generative models more robust and generalizable, and will be important for any applications of deep generative modeling to perform physical inference.

This work focuses on optimizing performance while using a fairly simple network architecture, but larger-scale applications may warrant more sophisticated schemes. If scaling up is done by stitching together sub-volumes, one needs to ensure continuity at the boundary of sub-volumes, and also ensure that spatial modes larger than the sub-volumes are properly modeled. One implementation uses super-resolution techniques and conditional neighbor information to generate large N-body volumes \citep{scale_up_gan1}. As described in Section \ref{sec:results_z0}, we experimented with simple extensions to our model, for example adding extra convolutional layers and expanding the latent space dimension, but did not see performance improvements. The issue of ``going deeper" is known to be delicate in the GAN literature \cite{biggan}, requiring advanced regularization techniques that we do not attempt in this work, but it may be possible to optimize neural network architecture in a more detailed fashion to capture the proper correlation structures seen in N-body simulations. 

Generally speaking, the applicability and practical utility of GAN models depend on the data quality of the training set and the interpolation accuracy of the generative model. While conditional GANs are only discussed briefly in this work (with redshift interpolation), we anticipate GANs will have significant utility as high-dimensional conditional emulators. Once larger, high-fidelity volumes can be generated, it may be possible to further condition the generator on cosmological parameters, for example $\Omega_m$ and $\sigma_8$. Many validation tests remain, but GANs could be integrated with larger, physically motivated simulation pipelines to bridge the gap between density fields and observables. With a sufficiently expressive and robust model, it may be possible to use GANs as signal priors that are fit directly to observational data through a sampling scheme or gradient based optimization.

\begin{acknowledgments}
R.M.F is supported by the California Institute of Technology. P.B. was supported by Jet Propulsion Laboratory, California Institute of Technology, under a contract with the National Aeronautics and Space Administration. Copyright 2020. All rights reserved. The authors would like to thank Tzu-Ching Chang, Olivier Dor\'e, Michael Albergo, Jeremy Bernstein and Yun-Ting Cheng for useful discussions.

The authors acknowledge the Texas Advanced Computing Center (TACC) at The University of Texas at Austin for providing {GPU cluster} resources that have contributed to the research results reported within this paper. \url{http://www.tacc.utexas.edu}. 

The authors also acknowledge the use of the following software for visualization and analysis: \texttt{h5py}; \texttt{Matplotlib}; \texttt{nbodykit} \citep{nbodykit}; \texttt{NumPy} \citep{numpy}; \texttt{powerbox} \citep{powerbox}; \texttt{Pylians}; \texttt{PyTorch}.
\end{acknowledgments}

\appendix
\section{Exploring the connection between discriminative and cosmological bias}\label{app:drs}

One challenge in using generative models to make predictions is that the statistics of generated samples are typically biased with respect to the data. For example, when generating cosmological volumes we never explicitly constrain the model to generate samples consistent up to the power spectrum, bispectrum, etc. Generative adversarial networks represent a framework in which an implicit likelihood is defined by a discriminator, which co-evolves with the generator -- there is no tailored loss function based on summary statistics. Even if one were used, inefficient training and/or insufficient model capacity often prohibit neural networks from fully capturing a given target distribution. While the working assumption is that a trained generator will produce samples with a distribution equal to that of the true distribution, certain samples may have artifacts or excess power that make them outliers with respect to $p_{data}$. On the other hand, because sample generation with GANs is fast, we are well justified to explore bias reduction techniques that may degrade the nominal computational performance. Most bias reduction techniques in the machine learning literature cite improvement according to neural network based metrics such as the Frechet Inception Distance (FID) \citep{fid} and the Inception Score (IS) \cite{inception_score}. However, it is unclear whether bias reduction with respect to the discriminator corresponds to bias reduction for cosmological metrics. 

One such bias reduction technique is discriminator rejection sampling (DRS) \cite{drs}. In standard rejection sampling, samples from a distribution $X$ with density $f(X)$ are drawn using samples from another distribution $Y$ with density $g(Y)$. This is done by computing $f(y)/Mg(y)$, where $M$ is some finite bound on the likelihood ratio $f(X)/g(X)$. The idea behind DRS is that, rather than perform post-processing to correct individual samples, one can correct \emph{ensemble based expectations} by training the discriminator network to learn the ratio $p_{data}(x)/p_g(x)$, where $p_{data}(x)$ and $p_g(x)$ are densities of the real and synthetic distributions, respectively.  For a fixed generator, minimizing the loss of the discriminator defined by a sigmoid and trained with cross-entropy loss yields
\begin{equation}
    D^*(x) = \frac{p_{data}(x)}{p_{data}(x)+p_{g}(x)}
\end{equation}
where $D^*$ is the optimal discriminator. One can also express the discriminator output in terms of the logit $\widetilde{D}(x)$ (again for a sigmoid activation):
\begin{align}
    D^*(x) = \frac{1}{1+e^{-\widetilde{D}^*(x)}} &= \frac{p_{data}(x)}{p_{data}(x)+p_g(x)}
\end{align}
Rearranging terms, the ratio $p_{data}(x)/p_g(x)$ can be expressed as
\begin{align}
    \frac{p_{data}(x)}{p_g(x)} &= e^{\widetilde{D}^*(x)}.
\end{align}
The likelihood ratio bound is set to $M = \max_x p_{data}(x)/p_g(x) = e^{\widetilde{D}^*(x^*)}$. In practice we cannot compute the true maximum, so instead we estimate a lower bound using a large number of samples. Finally, the acceptance probability for rejection sampling can be written as $\alpha = e^{\widetilde{D}^*(x)-\widetilde{D}_M^*}$. 

There are several caveats that degrade discriminator rejection sampling in practice. For one, the acceptance probability we estimate is only approximate to the true density ratio, since we cannot directly optimize over density functions. Furthermore, if the supports of $p_{data}$ and $p_g$ have a small intersection, then discriminator rejection sampling may yield vanishingly small acceptance probabilities. We counteract excessively small acceptance probabilities by modifying the logit $F(x)$ of the final sigmoid activation that defines the acceptance probability:
\begin{equation}
    \frac{1}{1+e^{-F(x)}} = e^{\widetilde{D}^*(x)-\widetilde{D}^*_M}.
\end{equation}
In the above equation, $F(x)$ can be computed from $\widetilde{D}^*(x)$ and $\widetilde{D}_M^*$, but in practice we include tunable parameters $\epsilon$ and $\gamma$:
\begin{equation}
  \hat{F}(x) = \widetilde{D}^*(x)-\widetilde{D}_M^* - \log\left(1-e^{\widetilde{D}^*(x) - \widetilde{D}^*_M - \epsilon}\right) - \gamma,
\end{equation}
After training the generator, we train the discriminator for an additional five epochs (with the generator fixed) such that it minimizes its loss, which plateaus around $10^{-6}$. Once this is done, we perform rejection sampling with hyperparameters $\epsilon=0.01$ and $\gamma = 0.0$, which gives an acceptance rate of $\sim 3\%$. Larger values of $\gamma$ lead to lower acceptance fractions, while more negative values yield the opposite. We calculate a lower bound for $\widetilde{D}^*(x)$ from 2000 generated samples.  

\begin{figure}
    \centering
    \includegraphics[width=\linewidth]{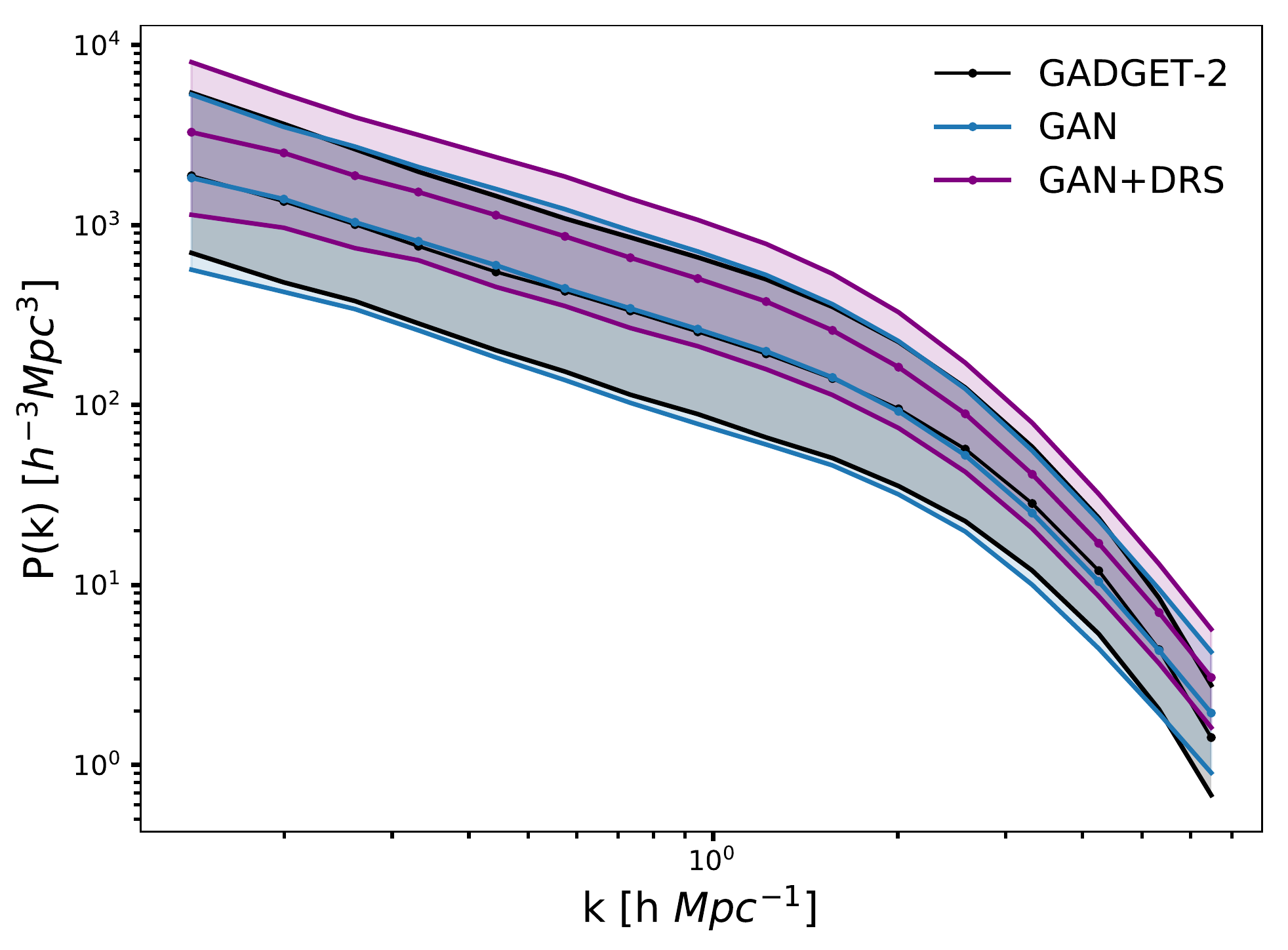}
    \caption{Comparison of matter power spectra between GADGET-2 sub-volumes at $z=0$ (black), samples drawn from a heavy-tailed GAN (blue) and samples obtained performing discriminator rejection sampling (DRS) on the heavy-tailed GAN (purple). Shaded regions bound the \nth{16} and \nth{84} percentiles of each ensemble and within each $k$ bin. These percentiles are calculated based on 2000 samples for each model and from the full GADGET-2 data set.}
    \label{fig:drs_pk}
    \centering
    \includegraphics[width=\linewidth]{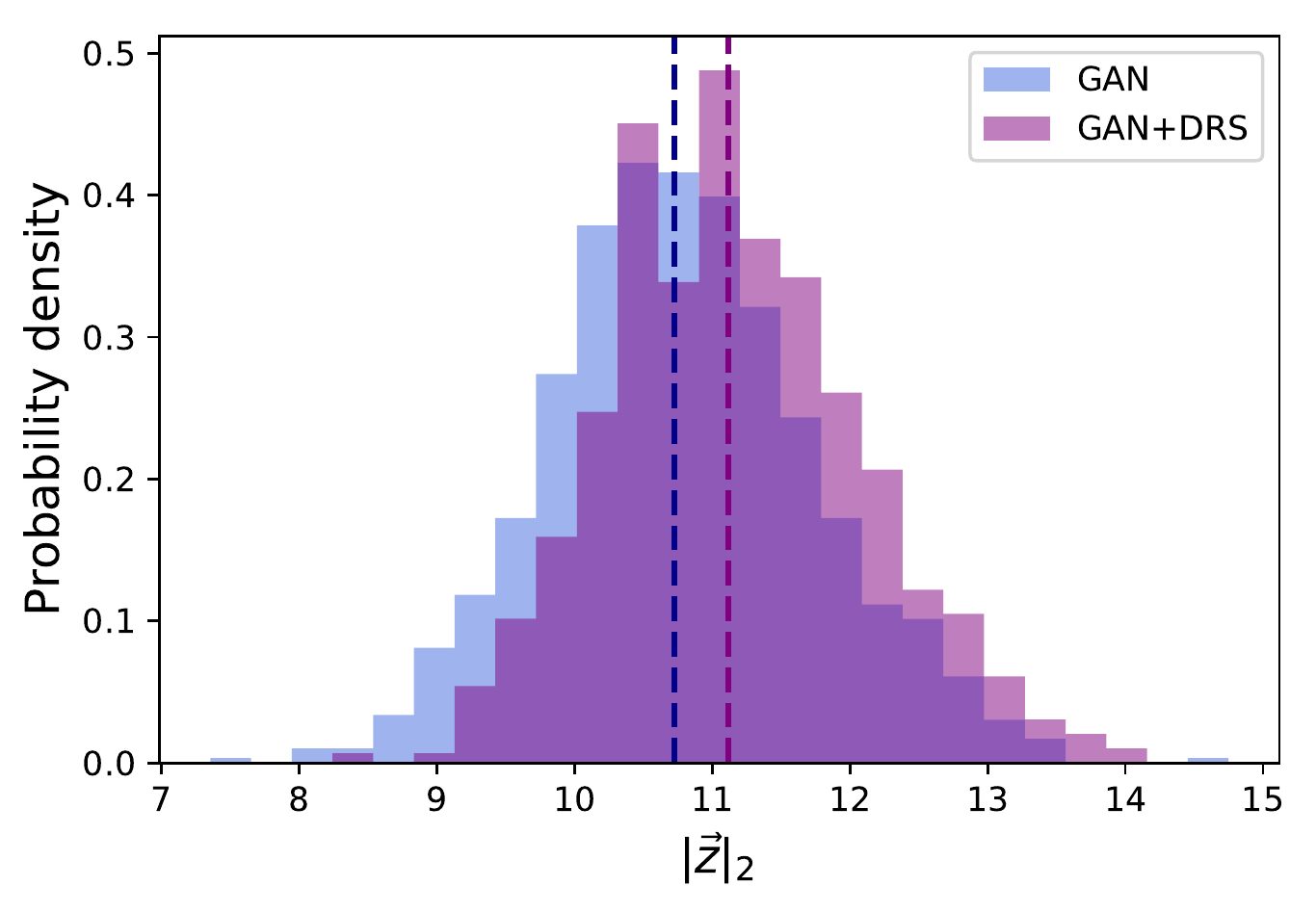}
    \caption{Comparison of $\ell_2$-norm distribution for latent vectors sampled by a heavy-tailed GAN (blue) and resampled latent vectors obtained after performing discriminator rejection sampling (purple). Dashed vertical lines indicate medians of the two distributions.}
    \label{fig:drs_l2z}
\end{figure}

Figure \ref{fig:drs_pk} shows a comparison of synthetic power spectra from a heavy-tailed GAN, both with and without DRS. The choice of a relatively unbiased model serves as a null test for DRS, i.e. the accuracy of generated power spectra should remain the same or improve if DRS is effective. To the contrary, we find that the trained discriminator preferentially accepts samples with higher power on average, ironically biasing our model more. We experimented with different values of $\gamma$ and found that when gamma was higher (i.e. when acceptance probabilities were lower), the accepted samples were more biased with respect to the mean power spectrum.

While DRS is evidently impractical for our purposes, it does provide a number of important insights to this work. For one, performing DRS demonstrates that our notion of cosmological bias (in one case, bias on the amplitude of the matter power spectrum) does not align with the more generic distributional ``bias" as determined by the discriminator. DRS also gives us insight into the latent space of the generator. Figure \ref{fig:drs_l2z} shows the $\ell_2$-norm distribution for a set of latent vectors drawn from the prior distribution, as well as latent vectors of samples accepted through DRS. One can see that DRS preferentially accepts samples drawn from latent vectors with larger magnitudes on average, while the scale dependence of the re-sampled power spectra remains intact. This suggests a relationship between the width of the re-sampled latent distribution and the power amplitude of samples. While these results are specific to the models trained in this work, DRS provides an opportunity to reason about how bias of the data distribution in discriminator-based, implicit likelihood methods compare with more physical notions of bias. 

\section{Scaling parameter}\label{app:scaling_param}
In Section \ref{sec:data_scaling}, we advocate for the use of a data scaling that preserves high-density features that have the most significant impact on our summary statistics. We motivate this by observing that when the scaling parameter $\kappa$ is small, the gradients in overdense regions (which can be very large in voxel space) get significantly suppressed in the scaled data. This is shown visually in Figure \ref{fig:data_scaling}, which has \eqref{scale_equation} plotted for different choices of $\kappa$. Capturing gradients in the high-density regime accurately with convolutions becomes more difficult, since small fluctuations in the final map may produce unrealistic values when scaled back to the original densities.

Another way to think about this problem is in terms of how a generic error gets propagated through the inverse data scaling. Recall the scaling used in this work:
\begin{equation}
    c = \frac{2\rho}{\rho+\kappa} - 1.    
\end{equation}
By computing the derivative of the inverse transformation, $d\rho/dc$, one can relate an error in $c$ to error in $\rho$:
\begin{equation}
    \delta \rho = \frac{2\kappa}{(1-c)^2}\delta c
\end{equation}
For a fixed density $\rho$, the ratio of scaled values for different values of $\kappa$ is
\begin{equation}
    \frac{c_1}{c_2} =\left(\frac{\rho-\kappa_1}{\rho-\kappa_2}\right)\left(\frac{\rho+\kappa_2}{\rho+\kappa_1}\right)
\end{equation}
For fixed $\rho$ and fixed $\delta c$,
\begin{align}
    \frac{\delta \rho_1(\kappa_1)}{\delta \rho_2(\kappa_2)} &= \frac{\frac{2\kappa_1}{(1-c_1)^2}\delta c}{\frac{2\kappa_2}{(1-c_2)^2}\delta c} \\
    &= \frac{\kappa_1}{\kappa_2}\left(\frac{1-c_2}{1-c_1}\right)^2
\end{align}
The last term in parentheses can be written as
\begin{equation}
    \frac{1-c_2}{1-c_1} = \frac{\kappa_2}{\kappa_1}\left(\frac{\rho+\kappa_1}{\rho+\kappa_2}\right)
\end{equation}
Finally,
\begin{align}
    \frac{\delta \rho_1(\kappa_1)}{\delta \rho_2(\kappa_2)} &= \frac{\kappa_1}{\kappa_2}\left(\frac{\kappa_2}{\kappa_1}\right)^2\left(\frac{\rho+\kappa_1}{\rho+\kappa_2}\right)^2 \\
    &= \left(\frac{\rho+\kappa_1}{\rho+\kappa_2}\right)^2\frac{\kappa_2}{\kappa_1}.
    \label{eqn:scaling_a}
\end{align}
As we are primarily concerned with the high-density regime, we can take the limit where $\rho \gg \kappa$ to conclude that $\delta \rho_1 / \delta \rho_2 \approx \kappa_2/\kappa_1$. This calculation is ill-posed in the sense that we have not identified specific, quantifiable uncertainties of the GAN model. Nonetheless, it demonstrates in a fairly generic setting how data scalings affect models with finite precision in reference to some target distribution. 

\begin{figure}
    \centering
    \includegraphics[width=\linewidth]{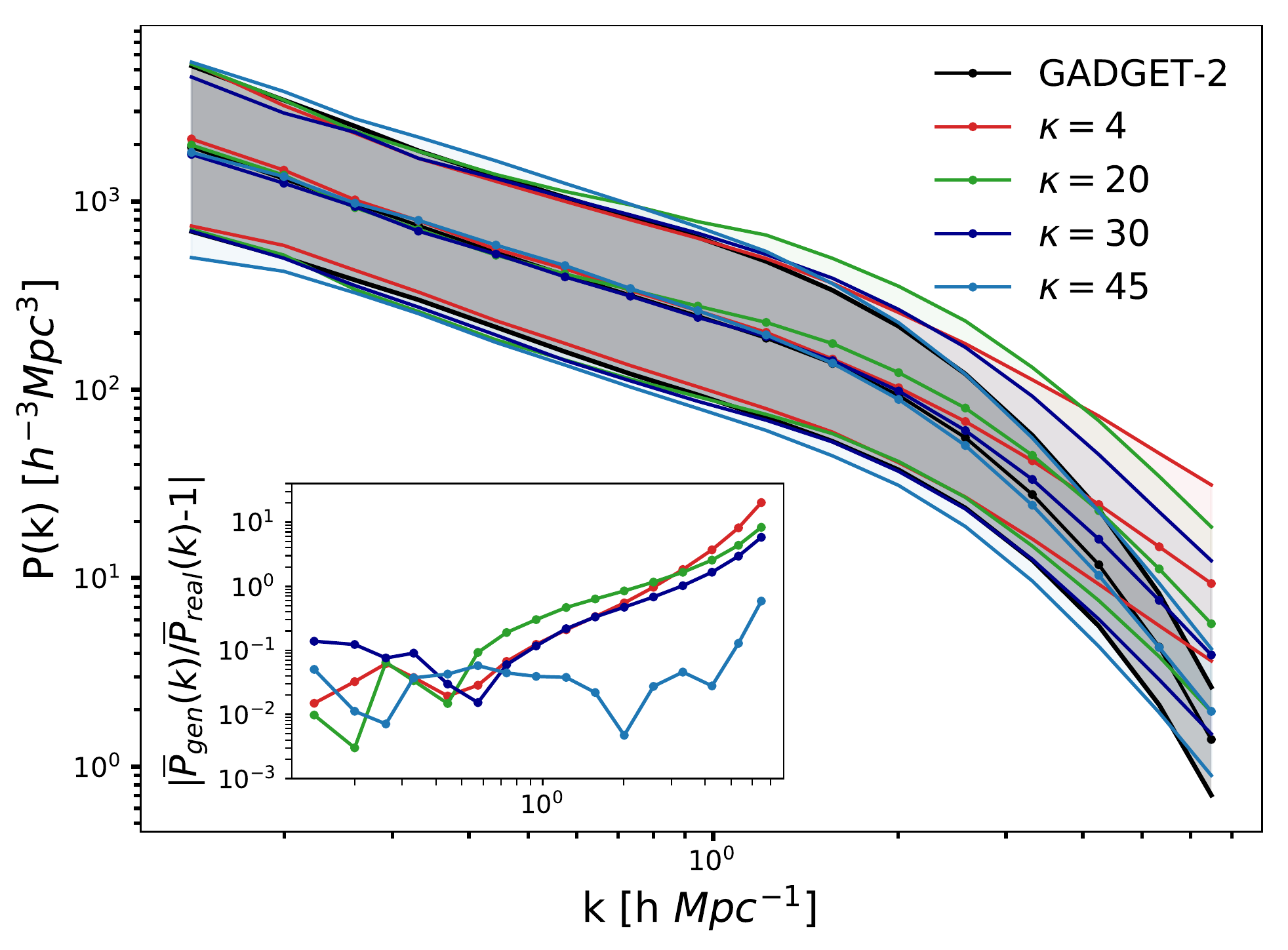}
    \caption{Comparison of real and synthetic power spectra, varying the data scaling parameter $\kappa$ from Equation \eqref{scale_equation} for each training run with a heavy-tailed GAN. Shaded regions bound the \nth{16} and \nth{84} percentiles of the ensemble within each $k$ bin, which are calculated based on 2000 samples for each model and from the full GADGET-2 data set. Higher values of $\kappa$ prioritize high-density features in the matter density field when the data are scaled to $[-1, 1)$.}
    \label{fig:change_kappa_pk}
\end{figure}

Figure \ref{fig:change_kappa_pk} shows the effect of $\kappa$ on the ability of our GANs to reproduce the power spectral distribution calculated from GADGET-2 N-body samples. While $\kappa=45$ produces a model that outperforms others on nearly all scales, the improvement is most significant for $k\gtrsim 1$, where the power spectrum accuracy is over a factor of ten better than models trained with smaller values of $\kappa$. Beyond $\kappa=45$, training was unstable and the generator was unable to learn a reliable representation of the data.

\section{Training data set reduction effect on performance} \label{app:training_data}
As data-driven models, generative adversarial networks require a large amount of training data to learn the underlying data distribution. Much like network architectures and training hyperparameters, the size and diversity of the training data set are important factors in the stability of GAN training. For a non-standard training set like our own, it is unclear how many training samples are needed to successfully train our GAN models. 

\begin{figure}[h]
    \centering
    \includegraphics[width=\linewidth]{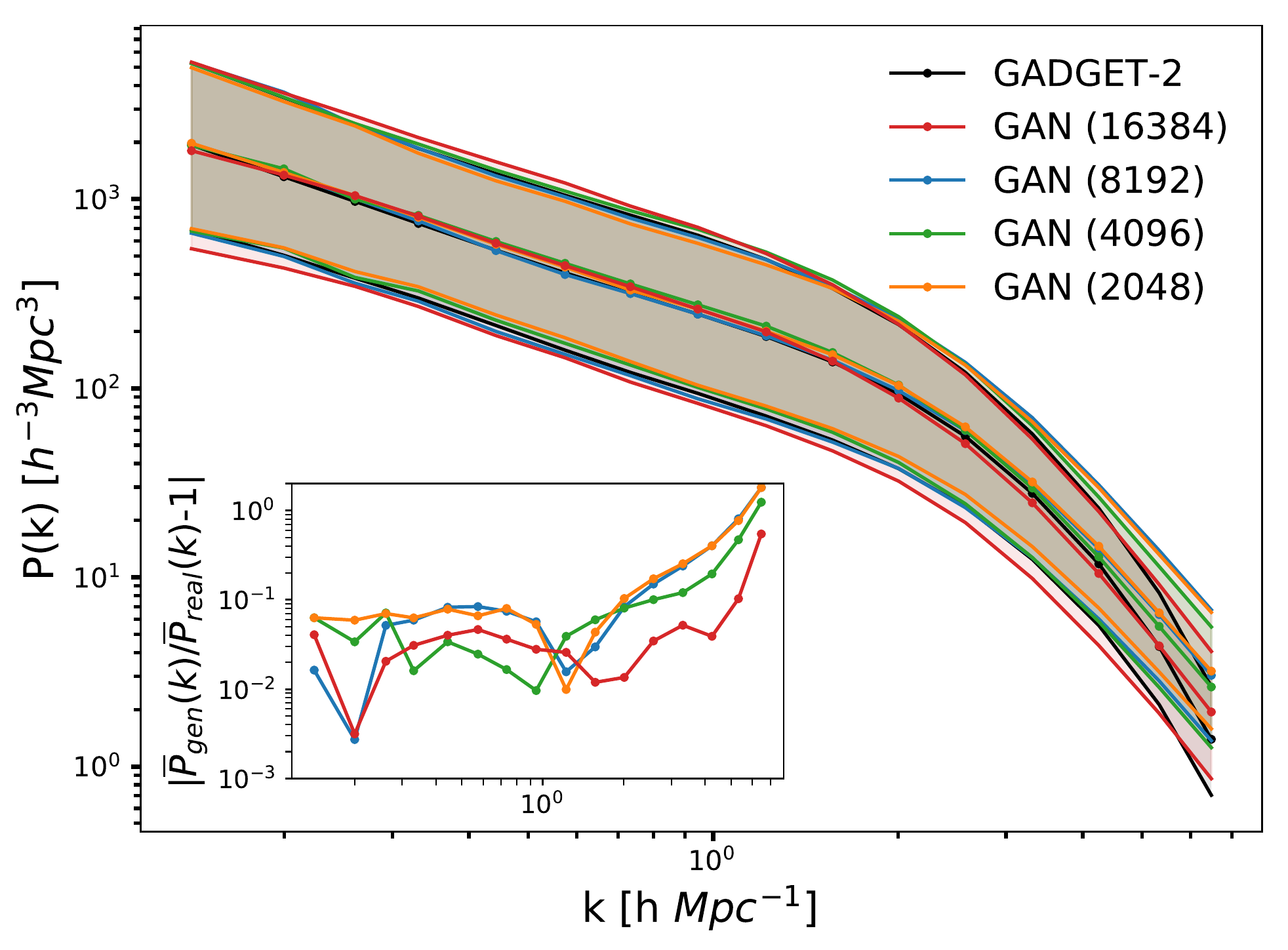}
    \includegraphics[width=\linewidth]{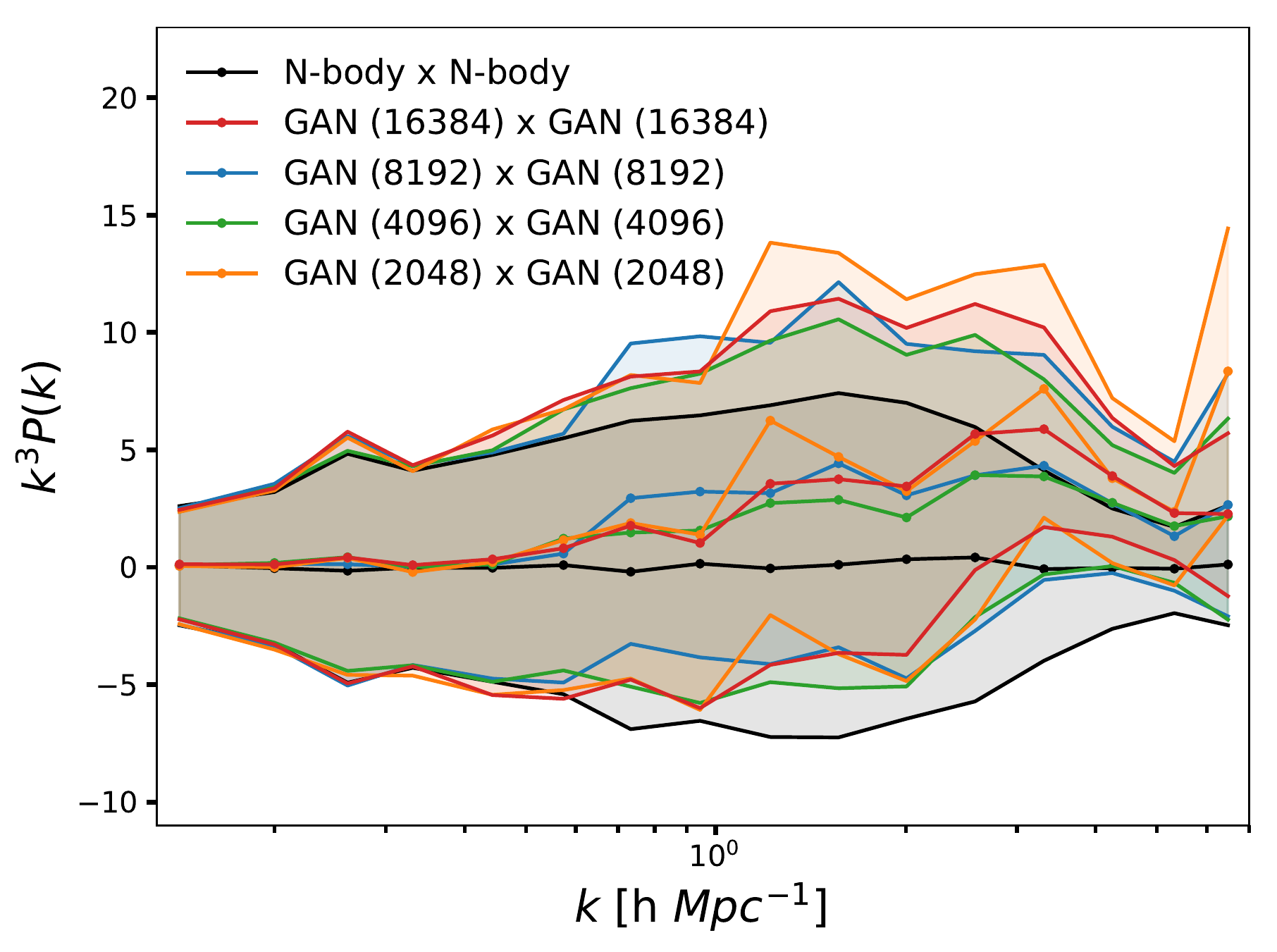}
    \caption{Comparison of synthetic power spectra (top) and cross-spectra (bottom) from GANs trained on the full training set (16384 samples), along with one-half (8192 samples), one-fourth (4096 samples) and one-eighth (2048 samples) of the full training set. Shaded regions bound the \nth{16} and \nth{84} percentiles of the ensemble within each $k$ bin, which are calculated based on 2000 samples for each model and from the full GADGET-2 data set.}
    \label{fig:reduced_data set_pk}
\end{figure}

To understand the effect of training set size on performance, we train three configurations of the same model using one-half, one-fourth and one-eighth of the full training set. We fix all hyperparameters in the GAN across runs. While some runs on smaller training sets required restarting the training to achieve stable gradient updates in the first epoch, all models were able to train successfully. Figure \ref{fig:reduced_data set_pk} shows power spectra (top) and cross-spectra (bottom) generated from the trained models. The power spectra generated from all three models are accurate to within $10\%$ up to $k=2$, but none of them performs as well in the highly non-linear regime as a model trained on the full data set. There is no clear relationship between the training set size and power spectrum accuracy among the four models, however the model trained on the least amount of data ($n_{train}=2048$) appears to have a larger cross-correlation on average between samples. 

\bibliography{apssamp}% Produces the bibliography via BibTeX.
\end{document}